\documentclass[preprintnumbers,superscriptaddress,showkeys,byrevtex]{revtex4}
\usepackage{bm}
\usepackage{orcidlink}
\usepackage{stackengine}
\usepackage{graphicx}
\usepackage{subcaption}
\usepackage{tikz-feynman}
\usepackage{pgfplots}
\usepgflibrary{plotmarks}
\begin{document}
\preprint{PKNU-NuHaTh-2025}
\title{Double $\phi$ Production in $\bar{p}p$ Reactions Near Threshold}
\author{Dayoung Lee\,\orcidlink{0009-0000-4828-3546}}
\affiliation{Department of Physics, Pukyong National University (PKNU), Busan 48513, Korea}
\author{Jung Keun Ahn\,\orcidlink{0000-0002-5795-2243}}
\address{Department of Physics, Korea University, Seoul 02841, Korea}
\author{Seung-il Nam$^*$\,\orcidlink{0000-0001-9603-9775}}
\email[]{sinam@pknu.ac.kr}
\affiliation{Department of Physics, Pukyong National University (PKNU), Busan 48513, Korea}
\date{\today}
\begin{abstract}
We employ an effective Lagrangian approach to investigate double $\phi$ production in the $\bar{p}p$ reaction near the threshold and describe a notable violation of the Okubo-Zweig-Iizuka (OZI) rule in this reaction process through hadronic degrees of freedom, using the currently available theoretical and experimental information. The ground-state nucleon and its $s$-wave resonances, i.e., $N^*(1535,1650,1895)$, are taken into account in the $t$- and $u$-channel tree-level diagrams. The pentaquark-like nucleon resonance $P_s(2071,3/2^-)$ is also considered. In the $s$-channel diagrams, scalar and tensor mesons are incorporated, specifically $f_0(2020,2100,2200)$ and $f_2(1950,2010,2150)$, respectively, along with the pseudoscalar meson $\eta(2225)$. Our calculations suggest that contributions from the $N$ and $N^*$ resonances significantly enhance the cross-section near the threshold. At the same time, those from the $f_0$ and $f_2$ mesons produce distinctive peak structures around $W=2.2$ GeV, qualitatively reproducing the JETSET experimental data. The openings of the $\bar{\Lambda}\Lambda$ and $\bar{\Sigma}\Sigma$ channels lead to the cusp structures, which can explain the nontrivial structures in the cross-section at $W\approx2M_{\Lambda}$ and $2M_{\Sigma}$ observed in the data. It turns out that contributions from the $\eta$ and $P_s$ resonances are almost negligible. Additionally, polarization observables such as the spin density matrix element (SDME) provide crucial insights into the individual hadronic processes involved in this reaction.
\end{abstract}
\maketitle
\section{Introduction}
In simple constituent quark models, the proton (antiproton) wave function contains only up and down quarks (antiquarks). Meanwhile, the $\phi$ meson is nearly a pure $\bar{s}{s}$ state. Consequently, the reaction $\bar{p}p\to\phi\phi$ may occur through two gluon emissions from the $\bar{q}q$ annihilation. All three valence quarks in the proton annihilate with the corresponding three antiquarks in the antiproton to create a purely gluonic state, from which $\phi\phi$ is formed. According to the Okubo-Zweig-Iizuka (OZI) rule, this process, with its disconnected quark lines,  should be strongly suppressed. On the other hand, the reaction $\bar{p}p\to\phi\phi$ may occur through a two-step process involving meson pairs, such as $\omega\omega$.  The $\omega\omega$ could be directly formed from the initial $\bar{p}p$ state, and the mixing of $\omega$ and $\phi$ mesons could lead to the creation of the $\phi\phi$ state. We can establish an upper limit for the total cross-section of the $\bar{p}p\to\phi\phi$ by comparing it to the total cross-section of the $\bar{p}p\to\omega\omega$, which is approximately $10$ nb and is expressed as $\sigma(\bar{p}p\to\phi\phi)=\tan^4\delta\cdot\sigma(\bar{p}p\to\omega\omega)\approx 10$ nb. This approximation is valid if independent OZI-violating couplings created both $\phi$ mesons. Here, the angle $\delta(=\Theta_i-\Theta)$ represents the difference between the ideal mixing angle $\Theta_i=35.3^\circ (\sin\Theta_i=1/\sqrt{3})$ and the mixing angle $\Theta$ between $(\phi,\omega)$ mesons  and the SU(3) states $(\omega_0,\omega_8)$. 

However, data from the JETSET experiment showed a significant violation of the OZI rule in the $\bar{p}p\to\phi\phi$ reaction~\cite{JETSET:1994evm,JETSET:1994fjp,JETSET:1998akg}. The measured cross-section of this reaction is $(2-4)\,\mu$b for incoming antiproton momenta ranging from $1.1$ to $2.0$ GeV$/c$. This value is two orders higher than the expected 10 nb, attributed to the $\phi$-$\omega$ mixing effect. A substantial OZI rule violation could indicate the presence of intriguing new physics. This violation can occur if a resonant gluonic state like a glueball or a four-quark state containing a significant $\bar{s}s$ admixture~\cite{Ke:2018evd,Lu:2019ira} contributes to the  $\bar{p}p\to\phi\phi$ reaction. According to lattice QCD results, the masses of the $2^{++}$ and $0^{-+}$ glueballs are predicted to be well above 2 GeV, around $(2.39\pm 0.12)$ and $(2.56\pm 0.12)$ GeV~\cite{Chen:2005mg}. However, some phenomenological model calculations point to the mass region near 2.0 GeV, which near-threshold $\phi\phi$ production experiments can access. A theoretical study using QCD sum rules estimates the masses of the $2^{++}$ and $0^{-+}$  glueballs to be $(2.0\pm 0.1)$ and  $(2.05\pm 0.19)$ GeV, respectively~\cite{Narison:1996fm}. In contrast, a recent QCD sum rule calculation predicts the masses of the $2^{++}$ and $0^{-+}$ glueballs to be $(1.86\pm 0.17)$ and $(2.17\pm 0.11)$ GeV~\cite{Chen:2021bck}.

It is suggested that strange quarks could be knocked off  directly from the $\bar{q}q$ sea of the proton and the antiproton to create a pair of 
$\phi$ mesons: $\phi\phi$. The strangeness content ($^1S_0~ \bar{s}s$) of the proton and antiproton might result in the production of $\phi\phi$ through a shake-out or rearrangement process~\cite{Ellis:1994ww}. Importantly, this process does not violate the OZI rule because it involves connected quark diagrams with higher Fock-space components in the nucleon wave function: $|p\rangle=x\sum^\infty_{X=0}|uudX\rangle + z\sum^\infty_{X=0}|uud\bar{s}sX\rangle, ~|x|^2+|z|^2=1,$ where $X$ stands for any number of glueons and light $\bar{q}q$ pairs. The upper limit for the total cross-section of the $\bar{p}p\to\phi\phi$ reaction is given by $\sigma(\bar{p}p\to\phi\phi)=(|z|/|x|)^4\cdot\sigma(\bar{p}p\to\omega\omega)\geq 250 {\rm~nb},$ which is larger than the value from the $\phi$-$\omega$ mixing effect, but still much smaller compared to the experimental data.   

The interaction between quarks, induced by instantons, could weaken the OZI suppression. A theoretical study~\cite{Kochelev:1995kc} demonstrates that the violation of the OZI rule in the $\bar{p}p$ annihilation is a nontrivial consequence of the complex structure of the QCD vacuum, which is associated with the existence of the instantons. On the other hand, the large cross-section for the $\bar{p}p\to\phi\phi$ reaction may be
explained by considering the hadronic rescattering mechanism. Each transition in the rescattering diagram is OZI-allowed. Lu {\it et al.} studied the role of a $\bar{K}K$ intermediate state in a triangle diagram in the $\bar{p}p\to\phi\phi$ reaction~\cite{Lu:1992xd}. The intermediate $\eta\eta$ can also contribute to the $\phi\phi$ production, as the $\eta$ contains $\bar{s}s$ content. In addition, the $\pi\pi\to \bar{K}K$ amplitude could make a sizable contribution. It is worth noting that the kernel $\bar{B}B\to\bar{m}m$ involving a baryon and antibaryon pair is possible. A full calculation involving all possible hadronic rescattering diagrams would be necessary to predict the detailed shape and magnitude of the observed spectrum.   

In the context of hadronic degrees of freedom, the $\bar{p}p\to\phi\phi$ reaction can be described using the meson and baryon exchange diagrams. Interestingly, the hadronic Yukawa vertices, such as the $\phi N N$ one, look like it violates the OZI rule, being indicated by its sizable coupling constant $g_{\phi NN}=-1.47$, which is about $10\,\%$ of the strangeless coupling $g_{\omega NN}=10.4$ in the Nijmegen potential models~\cite{Stoks:1999bz}. This observation can be understood by the fact that the sizable coupling originates from the unitarized vector-meson--baryon coupled-channel amplitude~\cite{Khemchandani:2011et}, whose $S=0$ channel contains $K^*\Lambda$, $K^*\Sigma$, $\omega N$, and $\rho N$ in addition to $\phi N$ interactions. Hence, the coupling provides finite strength beyond the OZI rule phenomenologically only with the hadronic degrees of freedom. Recent theoretical calculations suggest that the $N^*(1535)$ exchange in the $t$-channel could play a significant role and provide an essential source for bypassing the OZI rule~\cite{Shi:2010un}. Additionally, a more recent theoretical calculation indicates that the inclusion of either $f_0$ or $f_2$ in the $s$-channel can effectively describe the bump structure near $W\approx 2.2$ GeV~\cite{Xie:2014tra,Xie:2007qt}. These two previous works included only the $N^\ast(1535)$ exchange. 

Hence, in the present work, we employ an effective Lagrangian approach at the tree-level Born approximation to study $\bar{p}p\to\phi\phi$ near the threshold and explore the notable violation of the OZI rule only through the \textit{hadronic} degrees of freedom. We will not consider the quark and gluon contributions to the data, such as the glueball possibly, although its presence in the present reaction process can be crucial. The ground-state nucleon and its $s$-wave resonances, i.e., $N^*(1535,1650,1895)$, are taken into account in the $t$- and $u$-channel tree-level diagrams. The pentaquark-like nucleon resonance $P_s(2071,3/2^-)$ is also considered. The scalar and tensor mesons are incorporated, specifically $f_0(2020,2100,2200)$ and $f_2(1950,2010,2150)$, respectively, along with the pseudoscalar meson $\eta(2225)$, in the $s$-channel diagrams. The numerical results show that the contributions from the $N$ and $N^*$ resonances significantly increase the cross-section near the threshold. At the same time, those from the $f_0$ and $f_2$ mesons produce distinctive peak structures around $W=2.2$ GeV, qualitatively reproducing the JETSET experimental data. 

The openings of the $\bar{\Lambda}\Lambda$  and $\bar{\Sigma}\Sigma$ channels lead to the cusp structures, which could explain the nontrivial structures in the cross-section at $W\approx 2M_{\Lambda}$ and $2M_\Sigma$ observed in the data. It turns out that contributions from the $\eta$ and $P_s$ resonances are almost negligible.  We determine polarization observables by considering the different helicities for the final $\phi\phi$ states, i.e., the spin density matrix element (SDME). Therefore, the polarization data provides new information relevant to evaluating the resonance couplings. These observables extend our capabilities to validate the mechanisms of the reaction models used in data analyses through a combined fit of unpolarized cross-sections and polarization measurements. If further experiments confirm a significant violation of the OZI rule, an amplitude analysis of spin-dependent observables will be necessary, for which this paper lays the groundwork.

The paper is organized as follows: In Sec. II, we describe the reaction model for double $\phi$ production in $\bar{p}p$ reactions near the threshold. In Sec. III, we present the numerical calculation results for the total and differential cross-sections. Section IV focuses on spin density matrix elements and spin correlations between two $\phi$ mesons. Finally, Sec. V summarizes our conclusions.

\section{Theoretical Framework}
\begin{figure}[h]
\includegraphics[width= 16cm]{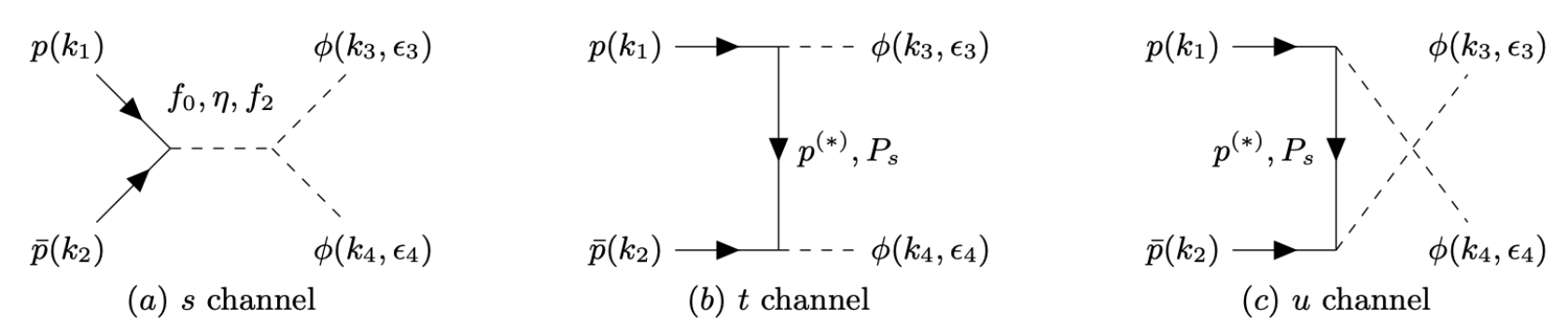}
\caption{The relevant Feynman diagrams illustrate the $(s,t,u)$-channel amplitudes for $\bar{p}p\to\phi\phi$. Solid lines represent (anti)proton and its resonances in the diagrams, while dashed lines represent scalar and tensor mesons. The four momenta ($k_i$) and polarizations ($\epsilon_i$) for the particles are also defined.}
\label{FIG0}
\end{figure}

\begin{figure}[h]
\includegraphics[width= 5.5cm]{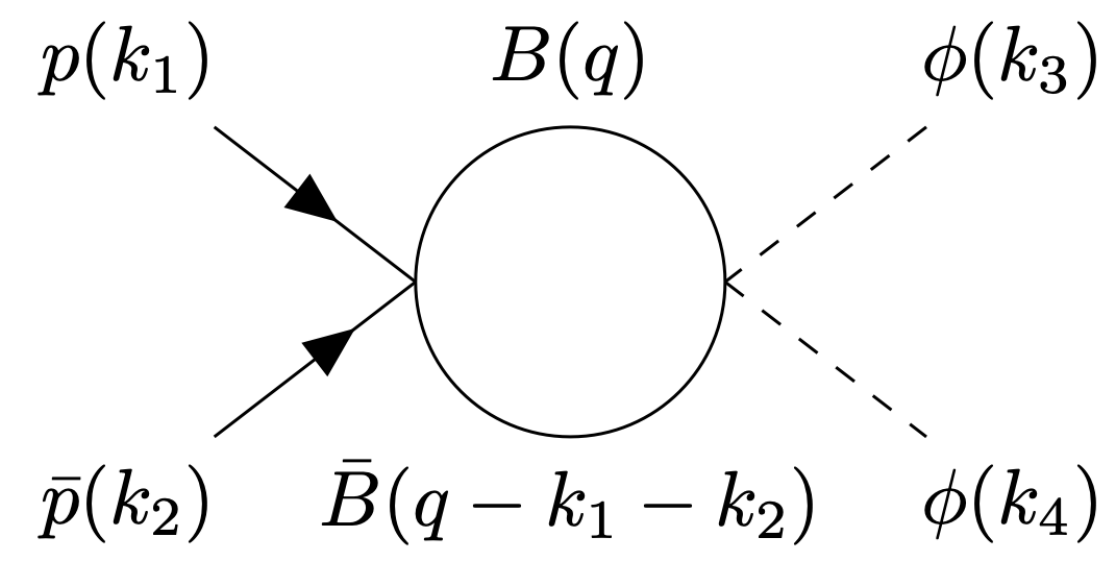}
\caption{A loop contribution for $\bar{B}B$ channel opening for $B=(\Lambda,\Sigma)$.}
\label{FIG1}
\end{figure}

This section briefly introduces the theoretical framework for studying the reaction process $\bar{p}p\to\phi\phi$. The relevant Feynman diagrams are provided in Fig.~\ref{FIG0}, along with the definition of the four momenta and polarization of the vector mesons. The effective Lagrangians for the Yukawa vertices are defined as follows:
\begin{eqnarray}
\mathcal L_{S NN} &=& g_{S NN}\bar{N}SN+{\rm h.c.},\,\,\,
\mathcal L_{SVV} =\frac{g_{S \phi\phi}}{ m_{\phi}}F_{\mu\nu}F^{\mu\nu}S,\,\,\,
\cr
\mathcal L_{PNN} &=& \frac{f_{PNN}}{M_P}\bar N\gamma_5(\rlap{/}\partial P)N,\,\,\,
\mathcal L_{PVV} = \frac{ig_{PVV}}{M_P}\epsilon_{\mu\nu\rho\sigma}F^{\mu\nu}_VF^{\rho\sigma}_VP,
\cr
\mathcal L_{T NN} &=&- \frac{ig_{TNN}}{M_N} 
\bar N(\gamma_{\mu}\partial_{\nu}+\gamma_{\nu}\partial_{\mu})N T^{\mu\nu}+{\rm h.c.},\,\,\,
\mathcal L_{TVV} =\frac{g_{TVV} }{2M_V}
\left[\frac{g_{\mu\nu}}{4}F_{\rho\sigma}F^{\rho\sigma}-{g^{\sigma\rho}}F_{\nu\rho}F_{\sigma\mu}\right]T^{\mu\nu}, 
\cr
\mathcal L_{VNN}&=&-g_{VNN}\bar{N}\left[\gamma_{\mu}-\frac{\kappa_{VNN}}{2M_N}\sigma_{\mu\nu}\partial^{\nu}\right]\Gamma_5NV^{\mu},\,\,\,
\mathcal L_{V NN'} = -\frac{ig_{V NN'}}{M_V}\bar{N}^{'\mu}\gamma^{\nu}(\partial_{\mu}V_{\nu}-\partial_{\nu}V_{\mu})\Gamma_5\gamma_5N+{\rm h.c.},
\end{eqnarray}
where $S$, $P$, $V$, $T$, and $(N,N')$ denote the scalar, pseudoscalar, vector, tensor, and nucleon fields for $J^P=(1/2^\pm,\,3/2^\pm)$, respectively, while the vector meson is given in the form of the antisymmetric field-strength tensor $F_{\mu\nu}=\partial_\mu V_\nu-\partial_\nu V_\mu$. Note that we employ the Lagrangians for the $SVV$ and $TVV$ interaction vertices, given in Refs.~\cite{Nam:2005jz,Katz:2005ir}. $\Gamma_5$ denotes $({\bf 1}_{4\times4},\gamma_5)$ for the parity-$(+,-)$ nucleon states. For the $N'$, we employed the Rarita-Schwinger formalism~\cite{Rarita:1941mf}.  By straightforwardly computing the invariant amplitudes using the interaction Lagrangians, we obtained the total amplitude, which is the sum of the following contributions illustrated in Fig.~\ref{FIG1}:
\begin{eqnarray}
i\mathcal M^s_{ S }&=&-\frac{2ig_{ S VV}g_{ S NN}}{M_{V}}
\frac{\bar{\nu}_{k_2}\left[(k_3\cdot k_4)(\epsilon_3\cdot\epsilon_4)-(k_3\cdot \epsilon_4)(\epsilon_3\cdot k_4)
\right]u_{k_1}}{s-M_{ S }^2+i\Gamma_{ S }M_{ S }}
\times F^{ S }_s,
\cr
i\mathcal M^s_P &=& \frac{ig_{PVV}f_{PNN}}{M_P^2}\frac{ \bar\nu_{k_2} \epsilon_{\mu\nu\rho\sigma}k_{3\mu}k_{4\nu}\epsilon_{3\rho}\epsilon_{4\sigma}\gamma_5\rlap{/}q_s u_{k_1}}{s-M_P^2+iM_P\Gamma_P}\times F^{P}_s,
\cr
i\mathcal{M}^s_{T }&=&\frac{ig_{T VV}g_{T  NN}}{2M_{V}M_N}\bar\nu_{k_2}
\left[ \frac{g_{\mu\nu}}{2}[(k_3\cdot k_4)(\epsilon_3\cdot\epsilon_4)-(k_3\cdot\epsilon_4)(\epsilon_3\cdot k_4)]
- g^{\sigma\rho} (k_{3\nu}\epsilon_{3\rho}-k_{3\rho}\epsilon_{3\nu})(k_{4\sigma}\epsilon_{4\mu}-k_{4\mu}\epsilon_{4\sigma}) \right] 
\cr
&\times&\left[\frac{G^{\mu\nu\alpha\beta}\left(\gamma_{\alpha} k_{1\beta}+\gamma_{\beta} k_{1\alpha}\right)}
{s-M_{T }^2+i\Gamma_{T }M_{T }}\right]u_{k_1}
\times F^{T }_s,
\cr
i\mathcal M^t_{N^{(*)}(1/2^\pm)}&=& -ig^2_{V NN^{(*)}}\bar{\nu}_{k_2}\Gamma_5
\left[\rlap{/}{\epsilon}_4-\frac{\kappa_{V NN^{(*)}}}{2M_{N^{(*)}}}\sigma_{\mu\nu}\epsilon_4^{\mu}k_4^{\nu}\right]
\left[\frac{\rlap{/}{q}_t+M_{N^{(*)}}}{t-M_{N^{(*)}}^2}\right]
\left[\rlap{/}{\epsilon}_3-\frac{\kappa_{V NN^{(*)}}}{2M_{N^{(*)}}}\sigma_{\mu\nu}\epsilon_3^{\mu}k_4^{\nu}\right]
\Gamma_5 u_{k_1}
\times F^{N^{(*)}}_c,
\cr
i\mathcal M^u_{N^{(*)}(1/2^\pm)}&=&i\mathcal M^t_{N^{(*)}}|_{3\leftrightarrow 4, t\to u}.
\cr
i\mathcal M^t_{N'^*(3/2^\pm)}&=&-\frac{g^2_{VNN'^*}}{M_V^2}\bar\nu_{k_2}\gamma_5\Gamma_5
(k_{4\mu}\epsilon_{4\nu}-k_{4\nu}\epsilon_{4\mu})\gamma^{\nu}
\left[g^{\mu\alpha}-\frac{1}{3}\gamma^\mu\gamma^\alpha-\frac{2}{3M^2_{N'^*}}q_t^\mu q_t^\alpha+\frac{q_t^\mu\gamma^\alpha+q^\alpha\gamma^\mu}{3M_{N'^*}}\right]
\cr
&\times&
\gamma^{\beta}(k_{3\alpha}\epsilon_{3\beta}-k_{3\beta}\epsilon_{3\alpha})\gamma_5\Gamma_5u_{k_1}\times F_c^{N'^*},
\cr
i\mathcal M^u_{N'^*(3/2^\pm)}&=&i\mathcal M^t_{N'^*}|_{3\leftrightarrow 4, t\to u}.
\label{eq:AMP}
\end{eqnarray}
Here, $q_{i\pm j}\equiv (k_i\pm k_j)$ and the Mandelstam variables are defined by $(s,t,u)=q^2_{s,t,u}$. We also define the tensor-meson propagator as follows:
\begin{eqnarray}
\label{eq:TENSOR}
G^{\mu\nu\alpha\beta}(s)&=&\frac{1}{2}(\bar g^{\mu\alpha}\bar g^{\nu\beta} + \bar g^{\mu\beta} \bar g^{\nu\alpha}) - \frac{1}{3}\bar g^{\mu\nu}\bar g^{\alpha\beta},\,\,\bar g^{\mu\nu}
= -g^{\mu\nu}+\frac{q_s^{\mu}q_s^{\nu}}{s}.
\end{eqnarray}
To incorporate the spatial extension of the hadrons, it becomes necessary to introduce phenomenological strong form factors to the amplitudes. In the present work, we use the following parameterization of the form factors:
\begin{eqnarray}
\label{eq:FF}
F^h_x\equiv F(x,M_h,\Lambda_h)=\frac{\Lambda^4_h}{\Lambda^4_h+(x-M_h^2)^2},
\end{eqnarray}
where $x=(s,t,u)$ and $h$ denotes the hadron species. As for the $t$- and $u$-channel amplitudes, we employ the following \textit{common} form factor, satisfying Lorentz invariance. Here, we consider similar vector-meson--baryon interactions as in the meson photoproduction and interchangeability of $t\leftrightarrow u$ ($k_3\leftrightarrow k_4$) in the $t$- and $u$-channel amplitudes~\cite{Davidson:2001rk,Nam:2005uq,Kim:2020wrd}:
\begin{eqnarray}
\label{eq:FC}
F^{N,N^*,N'^*}_c(t,u)=1-\left[1-F^{N,N^*,N'^*}_t(t)\right]\left[1-F^{N,N^*,N'^*}_u(u)\right].
\end{eqnarray}
 Fitting available experimental data will determine the cutoff mass $\Lambda_h$ later in Sec. III.

In $\bar{p}p$ scattering, $\bar{B}B$ channels can open and decay into two $\phi$ mesons. Hence, in the energy region from threshold to $W=2.5$ GeV, a cusp corresponding to the $\bar{\Lambda}\Lambda$ and $\bar{\Sigma}\Sigma$ loops can appear at $W=2M_\Lambda$ and $2M_\Sigma$, respectively. To describe the cusp effectively, we consider the one-loop diagram as depicted in Fig.~\ref{FIG1}. For those Yukawa interaction vertices shown in Fig.~\ref{FIG1}., we define the following point-interaction Lagrangians to simplify the problem:
\begin{eqnarray}
\label{eq:LL}
\mathcal{L}_{4B}=\frac{g_{4B}}{M^2_N}\bar{B}\bar{B}B'B',\,\,\,\mathcal{L}_{VBVB}=\frac{g_{VBVB}}{M^3_N}\bar{B}F_{\mu\nu}F^{\mu\nu}B,
\end{eqnarray}
where $B=(\Lambda,\Sigma)$. The unknown couplings $g_{4B}$ and $g_{VBVB}$ will be taken as free parameters here. Straightforwardly, the amplitude for the loop diagram can be computed as follows:
\begin{eqnarray}
\label{eq:LLL}
i\mathcal{M}_{\bar{B}B}&=&-g_{\bar{B}B}\bar{\nu}\left[(k_3\cdot k_4)(\epsilon_3\cdot\epsilon_4)
-(k_3\cdot \epsilon_4)(\epsilon_3\cdot k_4)\right]u\times F_\mathrm{loop}\times
\hskip-0.1cm \underbrace{\int\frac{d^4p}{(2\pi)^4}
\frac{{\rm{Tr}}[(\rlap{/}{p}+M_B)(\rlap{/}{p}+\rlap{/}{q}_{1+2}+M_B)]}
{[p^2-M^2_B][(p+q_{1+2})^2-M^2_B]}}_{G_{\bar{B}B}(s)},
\end{eqnarray}
where the reduced coupling reads $g_{\bar{B}B}\equiv g_{4B}g_{VBVB}/M^5_N$. The integral representing the ${\bar{B}B}$ loop, with cutoff regularization, is given by:
\begin{eqnarray}
\label{eq:GFUN}
G_{\bar{B}B}(s)&=&4i\int^1_0dx\left[I^{(2)}_{\bar{B}B}-\Delta I^{(0)}_{\bar{B}B}\right]
=-\frac{i}{4\pi^2}\int^1_0dx\,\Delta\ln\frac{\Delta_\mu}{\Delta},
\end{eqnarray}
where $x$ indicates the Feynman-parameterization variable and
\begin{eqnarray}
\label{eq:}
I^{(0)}_{\bar{B}B}(x)-\frac{1}{16\pi^2}\ln\frac{\Delta_\mu}{\Delta},\,\,\,\,
I^{(2)}_{\bar{B}B}(x)=-\frac{\Delta}{8\pi^2}\ln\frac{\Delta_\mu}{\Delta}.
\end{eqnarray}
Here, $\Delta=-x(1-x)s+M^2_B$ and $\Delta_\mu=-x(1-x)s+\mu^2$, in which $\mu$ stands for a cutoff mass, corresponding to the size of two baryon masses $\mu\approx2M_B$. To prevent the unphysical increase of $i\mathcal{M}_{\bar{B}B}$ caused by the terms involving $k_{3,4}$, we multiply the form factor $F_\mathrm{loop}=F^{N}_s(s,M_N,\Lambda_\mathrm{loop})$ to $i\mathcal{M}_{\bar{B}B}$. For simplicity, we consider the isospin averaged loop integral for $\Sigma^{0,\pm}$.
 
In interpreting the reaction mechanism of the double $\phi$ production process, the spin-density matrix element (SDME) is one of the useful observables. For the current reaction process, there are nine independent SDMEs, considering the two $\phi$-meson helicities, defined similarly in the previous study~\cite{Kim:2020wrd}. The $0$-th element of the SDME for the $\phi$-meson with $k_3$ ($\phi_3$) reads:
\begin{eqnarray}
\label{eq:SDME}
\rho^0_{\lambda_{\phi_3}\lambda^\prime_{\phi_3}}&=&
\frac{1}{2N_T}
\sum_{\lambda_{\bar{p}}}
\sum_{\lambda_p}
\sum_{\lambda_{\phi_4}=\pm1}
\mathcal{M}_{\lambda_{\bar{p}}\lambda_p\lambda_{\phi_3}\lambda_{\phi_4}}
\mathcal{M}^*_{\lambda_{\bar{p}}\lambda_p\lambda^\prime_{\phi_3}\lambda_{\phi_4}},\,\,\,
N_T =
\frac{1}{2}\sum_{\lambda_{\bar{p}}}
\sum_{\lambda_p}
\sum_{\lambda_{\phi_3}}
\sum_{\lambda_{\phi_4}=\pm1}
|\mathcal{M}_{\lambda_{\bar{p}}\lambda_p\lambda_{\phi_3}\lambda_{\phi_4}}|^2,
\cr
\rho^4_{\lambda_{\phi_3}\lambda^\prime_{\phi_3}}&=&
\frac{1}{N_L}
\sum_{\lambda_{\bar{p}}}
\sum_{\lambda_p}
\mathcal{M}_{\lambda_{\bar{p}}\lambda_p\lambda_{\phi_3}0}
\mathcal{M}^*_{\lambda_{\bar{p}}\lambda_p\lambda^\prime_{\phi_3}0},\,\,\,
N_L = \sum_{\lambda_{\bar{p}}}
\sum_{\lambda_p}
\sum_{\lambda_{\phi_3}}
|\mathcal{M}_{\lambda_{\bar{p}}\lambda_p\lambda_{\phi_3}0}|^2,
\end{eqnarray}
In this context, $\lambda_h$ represents the helicity of particle $h$ in a specific kinematic frame. We can obtain the SDMEs for the $\phi$-meson with $k_4$ ($\phi_4$) by simply swapping the subscript indices as $3\leftrightarrow4$ in Eq.~(\ref{eq:SDME}). To compare the SDMEs with experimental data, we need to boost the kinematic frame used for the theoretical computation to the $\phi$-meson rest frame. This involves using different spin-quantization axes, such as the helicity, Adair, and Gottfried-Jackson (GJ) frames, as defined in Ref.~\cite{Kim:2020wrd}.
\section{Numerical results and Discussions}
In this section, we present the numerical results along with their corresponding discussions. We examine the relevant mesons contributing to the $s$-channel in the current reaction process. For the scalar and tensor mesons near the threshold, we select $f_0(2020, 2100, 2200)$ and $f_2(1950, 2010, 2150)$, respectively. Based on experimental data from the Particle Data Group (2022) \cite{ParticleDataGroup:2022pth}, the pseudoscalar meson $\eta(2225)$ is known to be strongly coupled to $\phi\phi$, so we include this meson in our calculations.

All the relevant numerical inputs for the mesons are listed in Table~\ref{TAB1}. Here, we show the reduced coupling constants $g_\Phi = g_{\Phi\phi\phi}g_{\Phi NN}$ for $\Phi = (S, P, T)$ to overcome theoretical uncertainties, since theoretical and experimental information for $g_{\Phi\phi\phi}$ $g_{\Phi NN}$ are scarce. Ref.~\cite{ParticleDataGroup:2022pth} provides the couplings for the $\eta$ meson as $g_{\eta(\phi\phi, NN)} = (-4.062, 0.5)$.
\begin{table}[h]
\begin{tabular}{ |c||c|c|c|c|c|c|} 
\hline
&$f_0(2020)$&$f_0(2100)$&$f_0(2200)$&$f_2(1950)$&$f_2(2010)$&$f_2(2150)$\\
\hline
 $M-i\Gamma/2$ [MeV]&$1982-i218$&$2095-i143.5$&$2187-i103.5$&$1936-i232$&$2011-i101$&$2157-i76$\\
 \hline
 $g_{(S,P,T)}$ & \multicolumn{3}{c|}{$0.115$} & \multicolumn{3}{c|}{$-0.1$}\\
\hline
\end{tabular}
\caption{Relevant meson coupling constants for the $s$-channel contributions.}
\label{TAB1}
\end{table}

Now, we are in a position to discuss the contributions of nucleon resonances in the $t$ and $u$ channels of our numerical calculations. Xie $et\ al.$~\cite{Xie:2014tra} highlighted the importance of the strangeness content within nucleons for reproducing data, such as the $N^*(1535, 1/2^-)$. Another study by Khemchandani $et\ al.$~\cite{Khemchandani:2013nma} used the coupled-channel method within the framework of chiral dynamics, specifically the chiral unitary model (ChUM), to demonstrate strong coupling of three $s$-wave nucleon resonances to the $\phi$-$N$ channel: $N^*(1535, 1/2^-)$, $N^*(1650, 1/2^-)$, and $N^*(1895, 1/2^-)$. The resulting couplings, denoted as $g_{\phi NN^*}$, are listed in Table~\ref{TAB2}.

Furthermore, as discussed in our previous work~\cite{Nam:2021ayk}, a possible pentaquark baryon, $P_s(2071, 3/2^-)$, is considered as a $K^*\Sigma$ bound state that decays into $\phi N$. Its coupling has been calculated using ChUM~\cite{Khemchandani:2011et} and is presented in the table. Note that we set the values of the tensor-interaction strength $\kappa_{N^*}$ to zero due to limited information, except for $\kappa_{\phi NN}$, which is fixed at $-1.65$~\cite{Kim:2021adl}. For simplicity in the computations, we use a single cutoff mass $\Lambda_{h,\mathrm{loop}} = (550,300)$ MeV for all hadronic form factors with $g_{\bar{\Lambda}\Lambda,\bar{\Sigma}\Sigma} = (2,1.2)$ by fitting available experimental data.

\begin{table}[h]
\begin{tabular}{|c||c|c|c|c|c|c|c|} 
\hline
&$N$&$N^*(1535,1/2^-)$& \multicolumn{2}{c|}{$N^*(1650,1/2^-)$} & \multicolumn{2}{c|}{$N^*(1895,1/2^-)$}&$P_s(2071,3/2^-)$\\
\hline
$M-i\Gamma/2$ [MeV]&$938-i0$&$1504-i55$&$1668-i28$&$1673-i67$&$1801-i96$&$1912-i54$&$2071-i7$\\
\hline
$g_{\phi NN^{(*)}}$&$-1.47$&$1.4+i2.2$&$4.1-i2.7$&$4.5+i5.2$&$2.1+i1.8$&$0.9-i0.2$&$0.14+i0.2$\\
\hline
 \end{tabular}
\caption{Relevant nucleon coupling constants for the $t$- and $u$-channel contributions~\cite{Khemchandani:2013nma,Khemchandani:2011et}.}
\label{TAB2}
\end{table}

\begin{figure}[t]
\begin{tabular}{cc}
\topinset{(a)}{\includegraphics[width= 8.5cm]{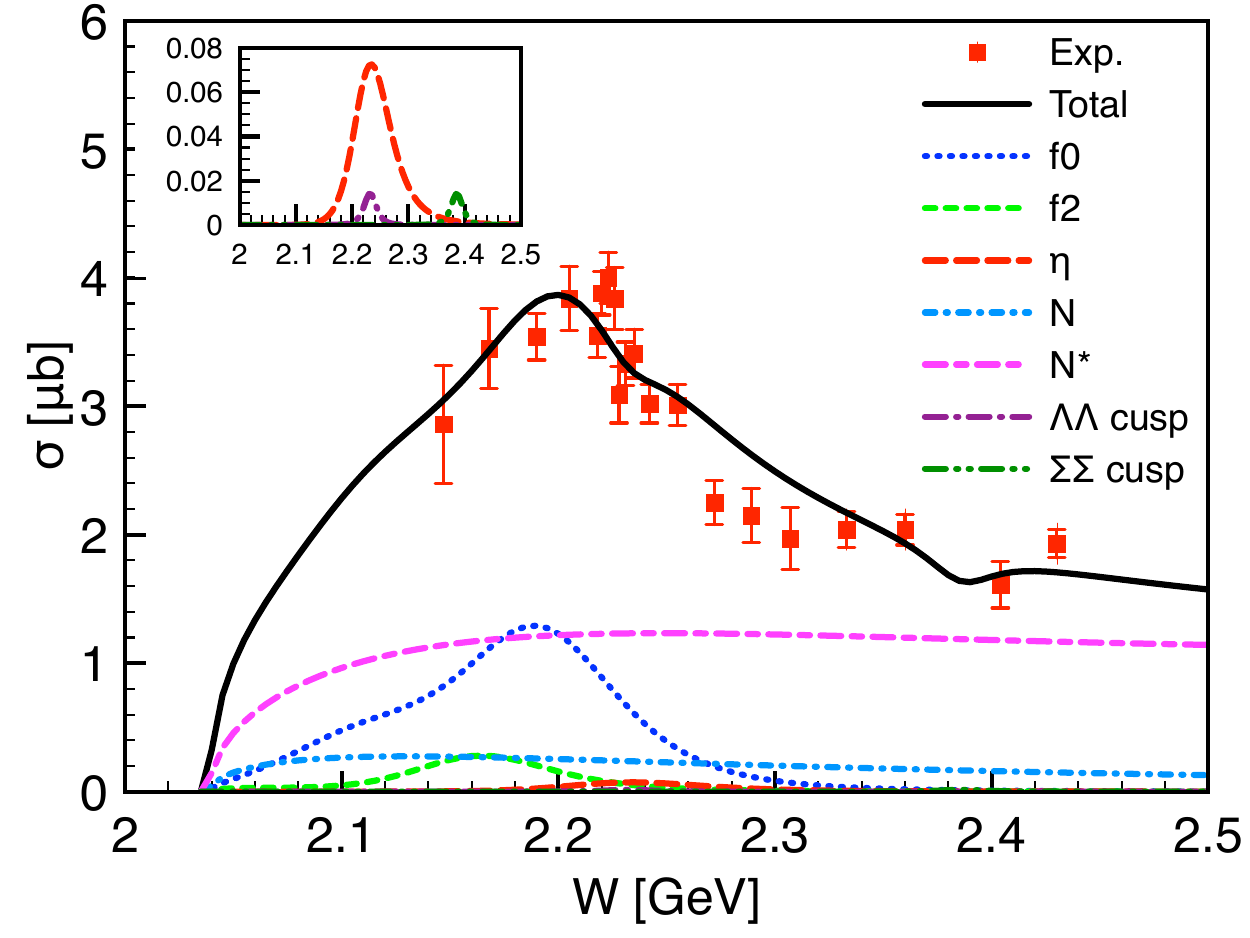}}{-0.2cm}{0.5cm}
\topinset{(b)}{\includegraphics[width= 8.5cm]{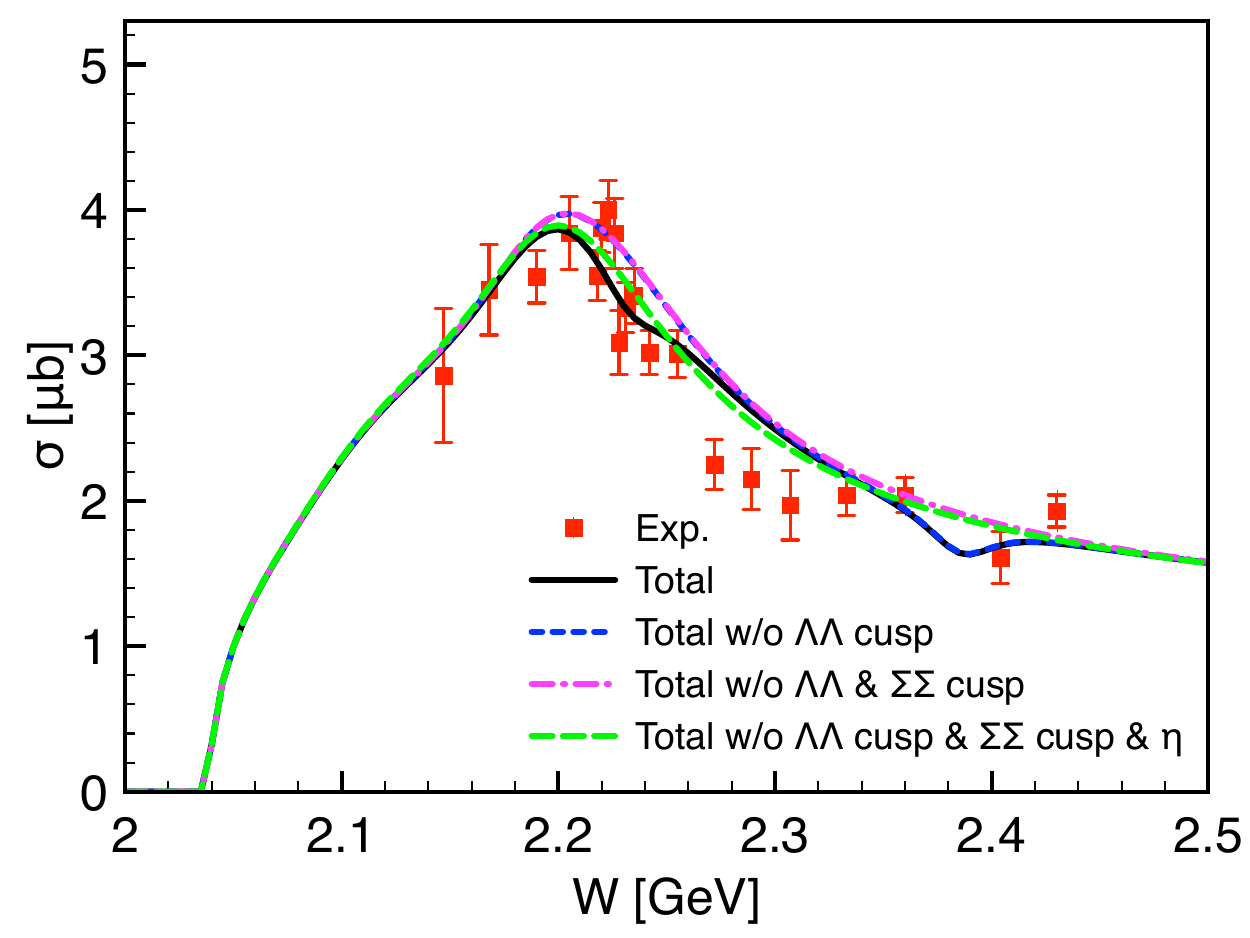}}{-0.2cm}{0.5cm}
\end{tabular}
\begin{tabular}{cc}
\topinset{(c)}{\includegraphics[width= 8.5cm]{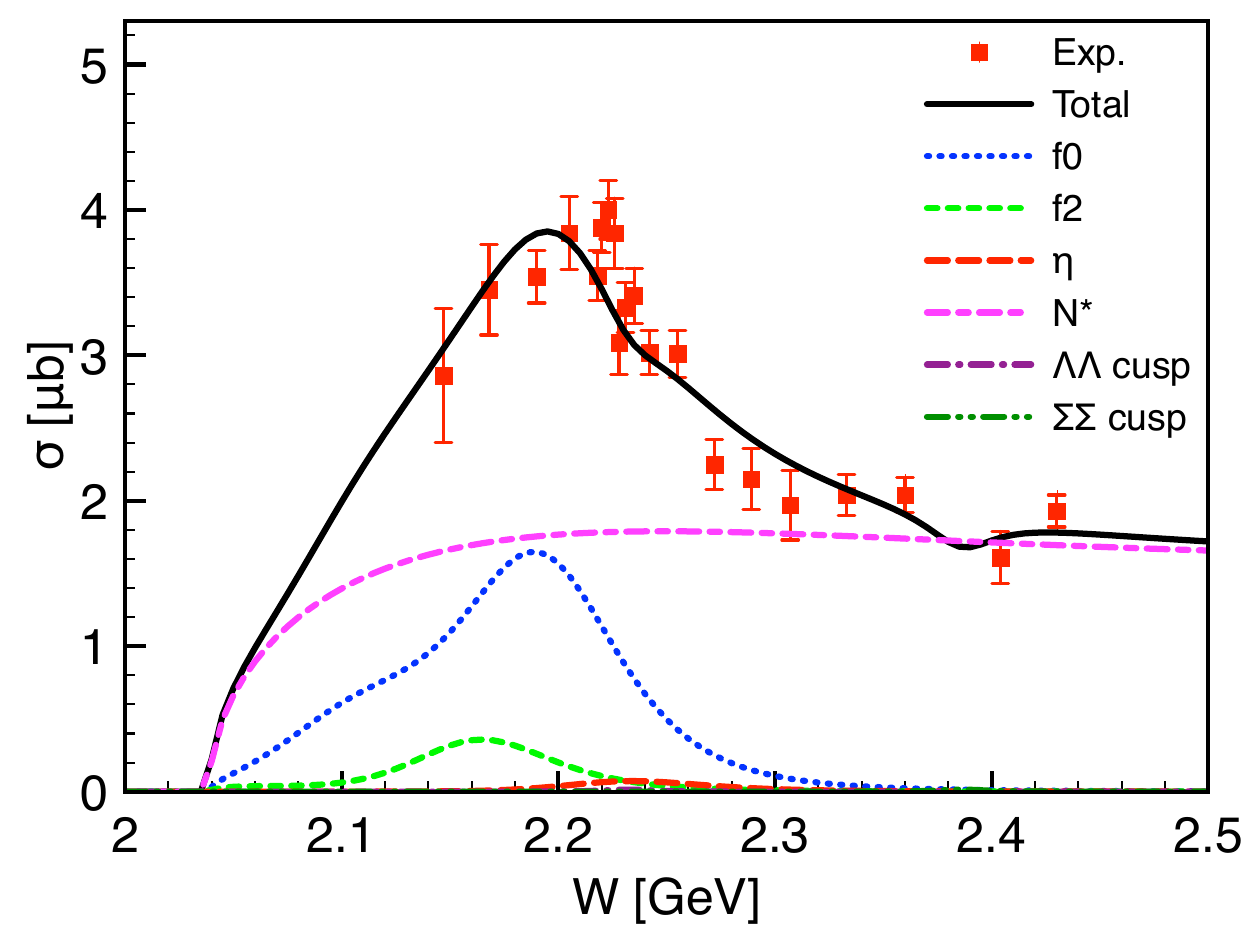}}{-0.2cm}{0.5cm}\,\,\,\,
\topinset{(d)}{\includegraphics[width= 7.5cm]{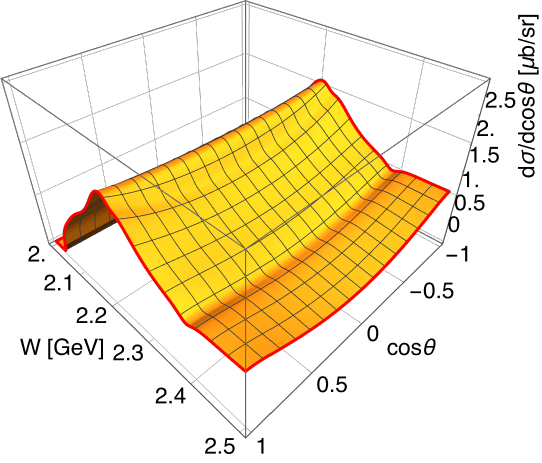}}{-0.2cm}{0.5cm}
\end{tabular}
\caption{(Color online) (a) Total cross-sections $\sigma\equiv\sigma_{\bar{p}p\to\phi\phi}$ as functions of $W$, showing each contribution separately. Experimental data are taken from Ref.~\cite{JETSET:1994evm,JETSET:1994fjp,JETSET:1998akg}. (b) Those with and without the cusp effect in addition to the $\eta$ contribution. (c) Those without the ground-state nucleon ($N$) contribution. (d) Angular-dependent differential cross-section $d\sigma/d\cos\theta$ as a function of $W$ and $\cos\theta$.}
\label{FIG2}
\end{figure}

In panel (a) of Fig.~\ref{FIG2}, we present the full calculations for the total cross-sections $\sigma \equiv \sigma_{\bar{p}p\to\phi\phi}$ as functions of the center-of-mass energy $W$, showing each contribution separately. The experimental data are taken from Ref.~\cite{JETSET:1994evm,JETSET:1994fjp,JETSET:1998akg}. 
The ground-state nucleon ($N$) contribution is significant near the threshold, exhibiting a shoulder-like structure, while the nucleon-resonance ($N^{*}$) contribution, including $P_s$, becomes stronger as W increases. We verified that the effect of the $P_s$ is much smaller than other nucleon resonances. As expected, the scalar and tensor mesons $f_{0,2}$ are responsible for the bump structure around $W=2.2$ GeV. Additionally, there is a small but finite contribution from the $\eta$ in the $s$ channel. Interestingly, the nontrivial structure around $W\approx2M_\Lambda$ and $2M_\Sigma$ are well reproduced by the interference between the cusp effects from the $\bar{\Lambda}\Lambda$- and $\bar{\Sigma}\Sigma$-loops and others. To clarify this observation, in panel (b), we show the total cross-section with and without the cusp effects. Although we try to explain the structure with the cusps from the $\bar{B}B$ channel opening here, we admit that interference with unknown mesonic resonances can also explain the structure. Moreover, the singularity from a complicated loop can be responsible for it as studied in different processes~\cite {Szczepaniak:2015eza}. We want to explore these possibilities in separate future works. We also test the impact of the $\eta$ contribution at $W \approx 2.25$ GeV, which fails to explain the nontrivial structure.

Following a similar approach to Refs.~\cite{Shi:2010un,Xie:2014tra,Xie:2007qt}, in panel (c), we attempt to reproduce the data without the ground-state nucleon contribution by modifying the cutoff masses for the form factors, providing a compatible result with the full calculation shown in panel (a). As expected, the absence of the $N$ contribution causes the shoulder-like structure near the threshold to disappear in panel (c). Additionally, the curve shows better behavior in the higher-energy region beyond $W = 2.4$ GeV compared to panel (a). We will further explore these two scenarios, $N + N^*$ (full) and $N^*$ only, in detail later. In panel (d), we present the numerical results for the angular-dependent differential cross-section $d\sigma/d\cos\theta$ as a function of $W$ and $\cos\theta$, where $\theta$ denotes the scattering angle of the $\phi_3$ in the center-of-mass system for the full calculation. As shown, the angular dependence is symmetric about $\theta = \pi/2$ since identical mesons are scattered in the final state. The cross-section decreases as $\theta$ approaches $\pi/2$ from $\theta=0$.

\begin{figure}[t]
\begin{tabular}{ccc}
\topinset{(a) $W=2.1$ GeV}{\includegraphics[width= 5.5cm]{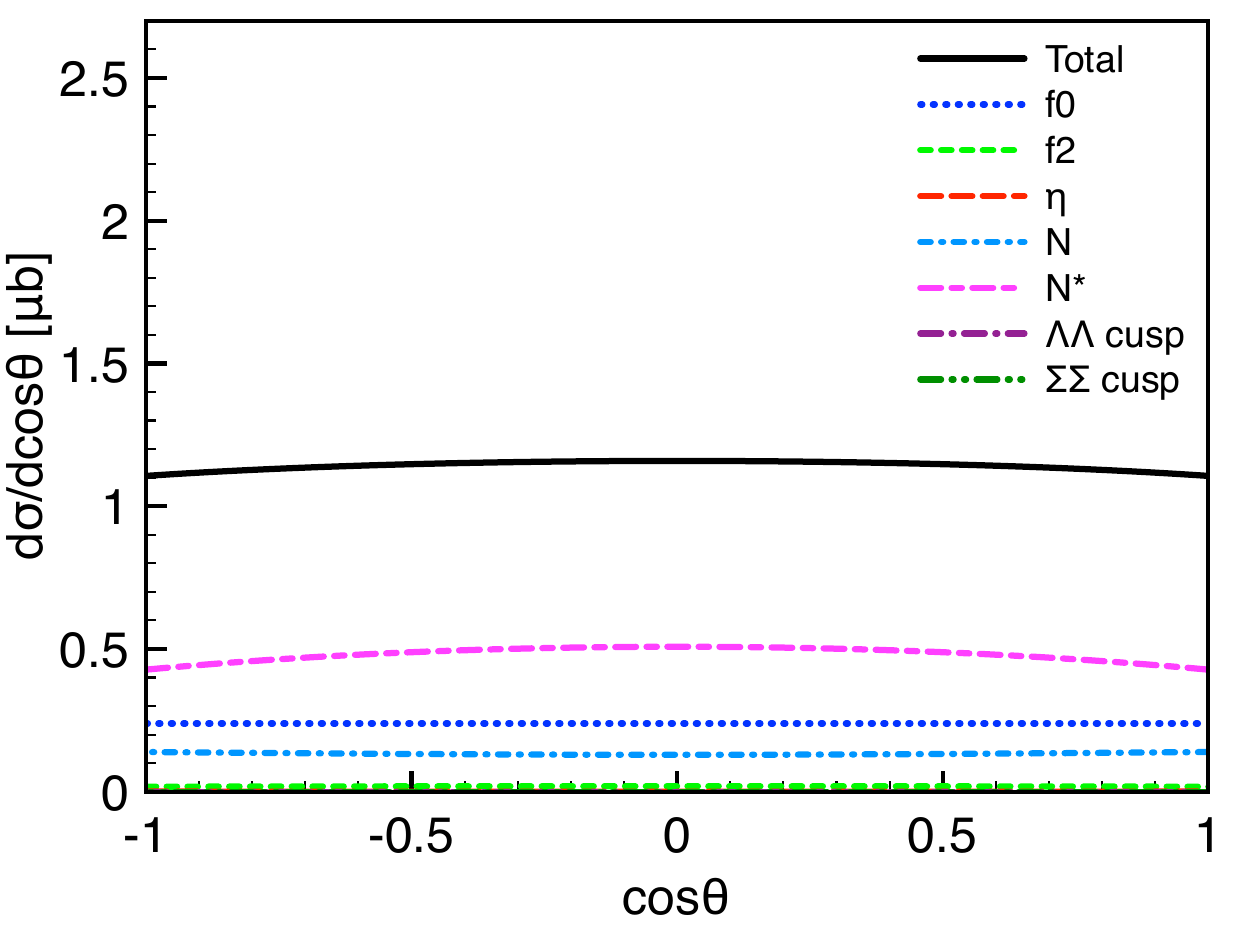}}{-0.3cm}{0.5cm}
\topinset{(b) $W=2.2$ GeV}{\includegraphics[width= 5.5cm]{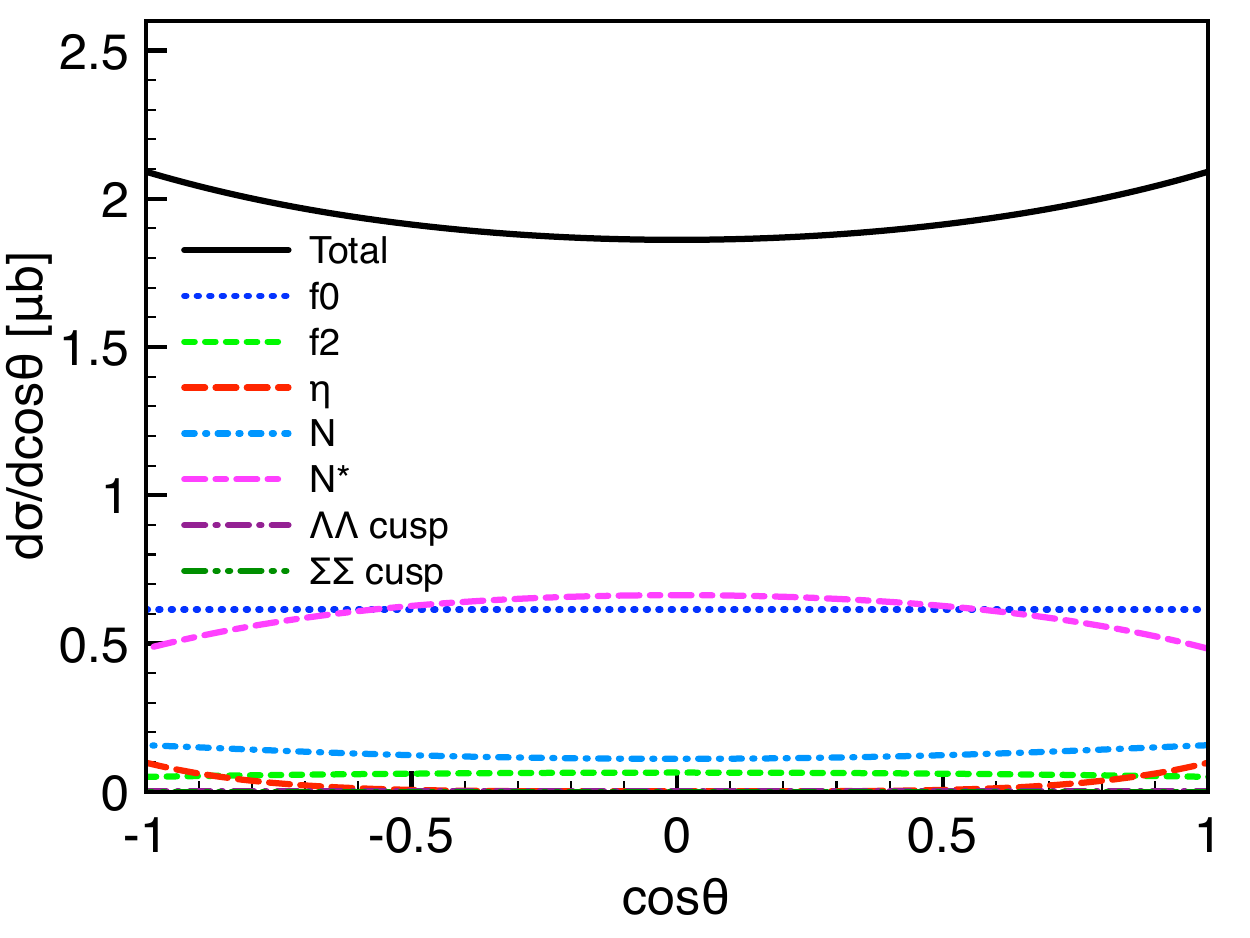}}{-0.3cm}{0.5cm}
\topinset{(c) $W=2.3$ GeV}{\includegraphics[width= 5.5cm]{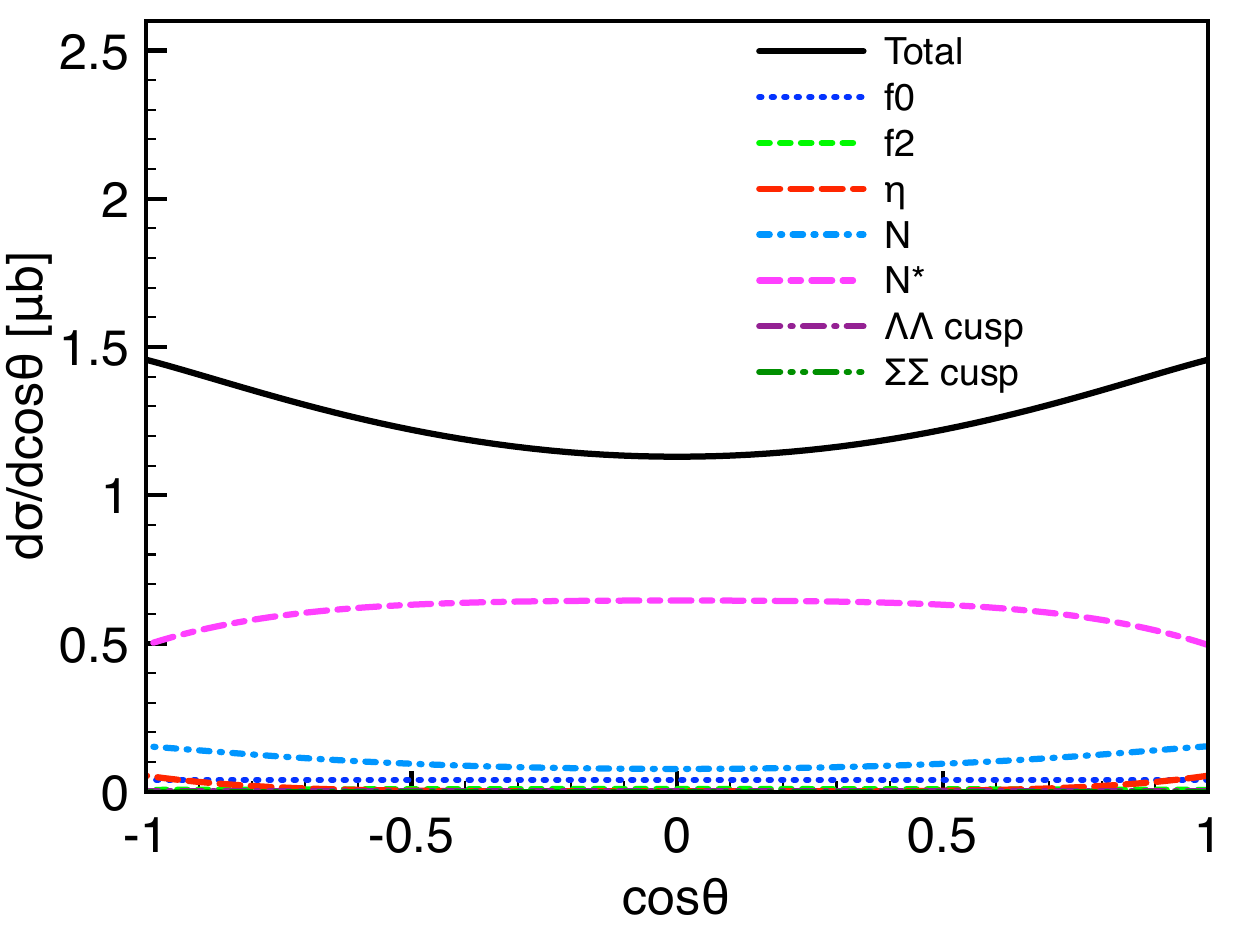}}{-0.3cm}{0.5cm}
\end{tabular}
\begin{tabular}{ccc}
\topinset{(d) $\theta=0$}{\includegraphics[width= 5.5cm]{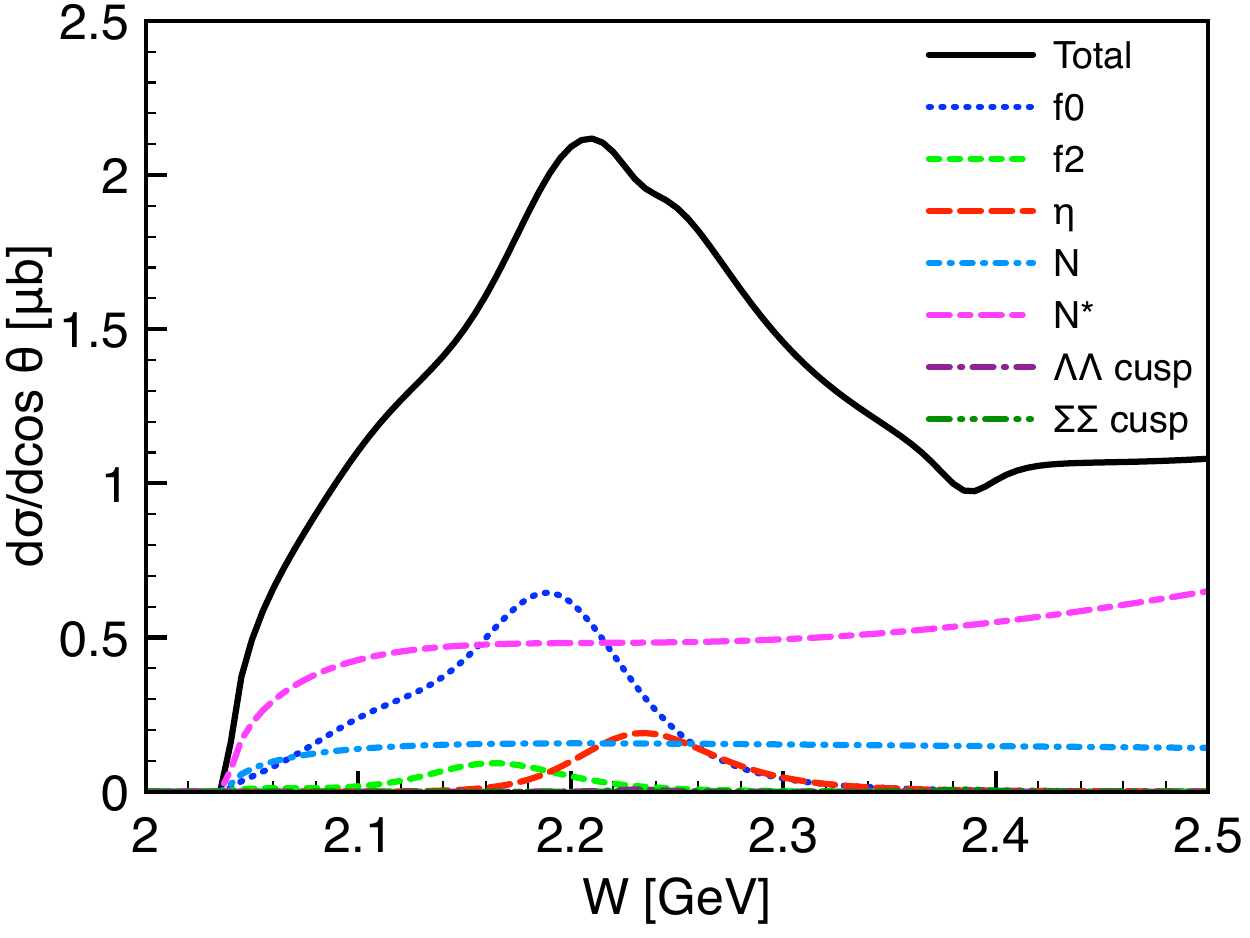}}{-0.3cm}{0.5cm}
\topinset{(e) $\theta=\pi/4$}{\includegraphics[width= 5.5cm]{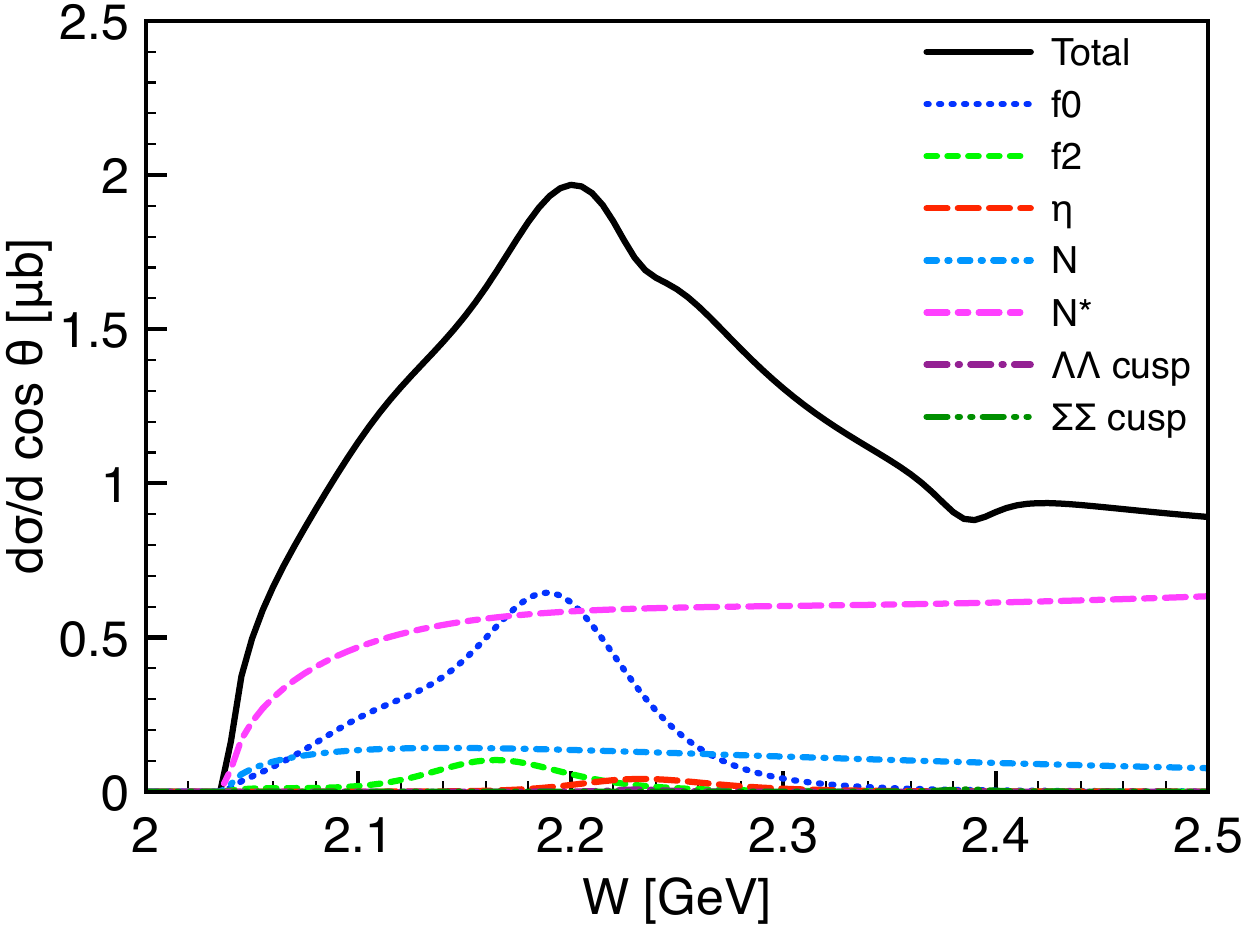}}{-0.3cm}{0.5cm}
\topinset{(f) $\theta=\pi/2$}{\includegraphics[width= 5.5cm]{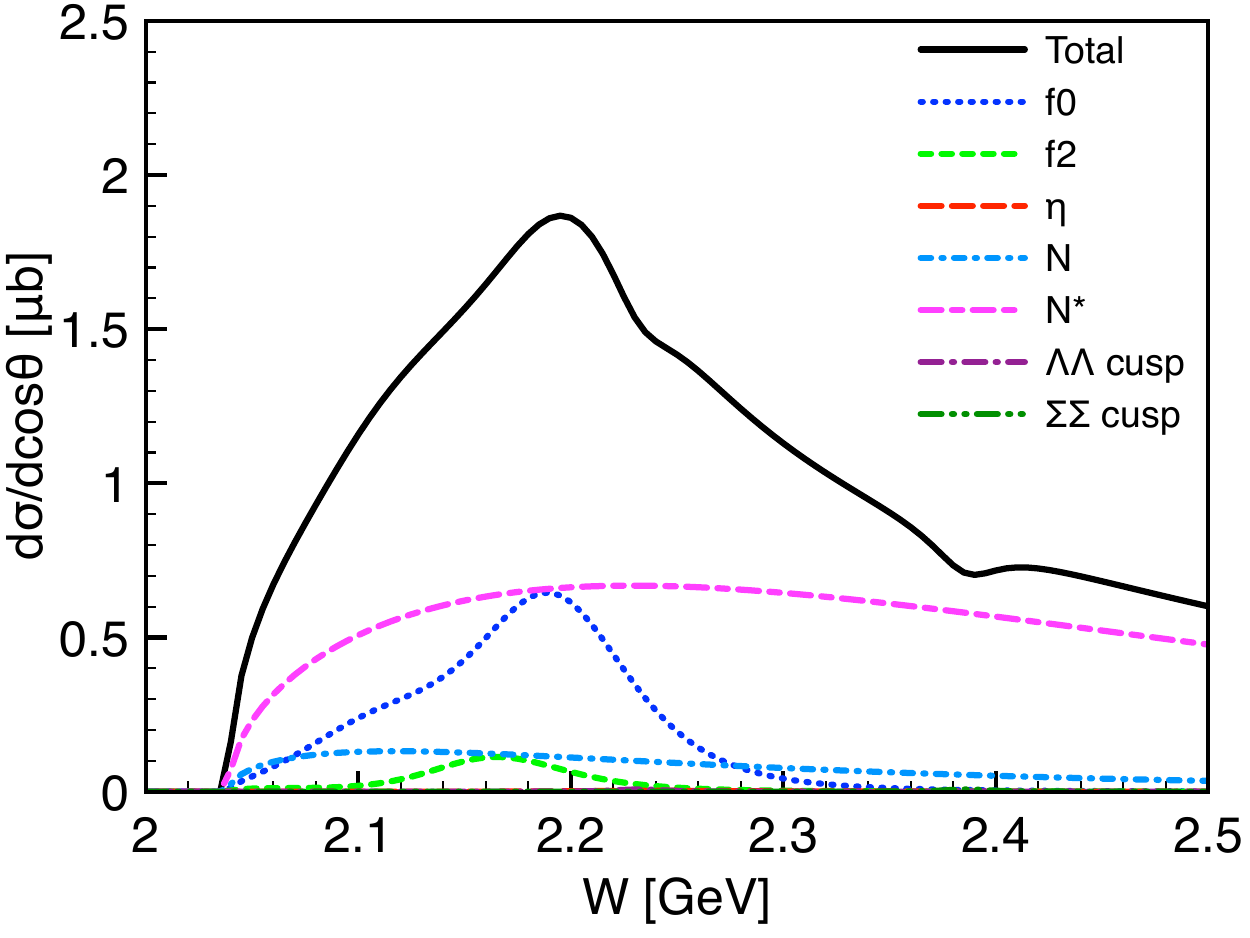}}{-0.3cm}{0.5cm}
\end{tabular}
\caption{(Color online) (a-c) Angular-dependent differential cross-sections $d\sigma/d\cos\theta$ as functions $\cos\theta$ at different $W$. (d-f) The same as functions of $W$ for the different angles.}
\label{FIG3}
\end{figure}

To better understand the angular dependence of the present reaction process, in panels (a - c) of Fig.~\ref{FIG3}, we show the full results for $d\sigma/d\cos\theta$ as a function of $\cos\theta$ at different energies $W = (2.1 - 2.3)$ GeV. We also display the separate contributions in the panels. Near the threshold at $W = 2.1$ GeV, all contributions are nearly flat, with the ground-state nucleon contributing the most to the cross-section. As the energy increases, the $N$ and $N^*$ contributions become more significant, leading to non-trivial angular dependences, such as convex and concave shapes. We also observe that the meson contributions are maximized around $W =2.2$ GeV. Note that the $N$ contribution primarily determines the total curves and their angular dependences. In panels (d - f), we present the full calculation results for $d\sigma/d\cos\theta$ as functions of $W$ for different angles $\theta = (0 - \pi/2)$. As already noted in panel (d) of Fig.~\ref{FIG2}, the magnitude of the cross-sections decreases mainly due to the diminishing $\eta$ and $N$ contributions as the angle increases.

\begin{figure}[b]
\begin{tabular}{cc}
\topinset{(a)}{\includegraphics[width= 8.5cm]{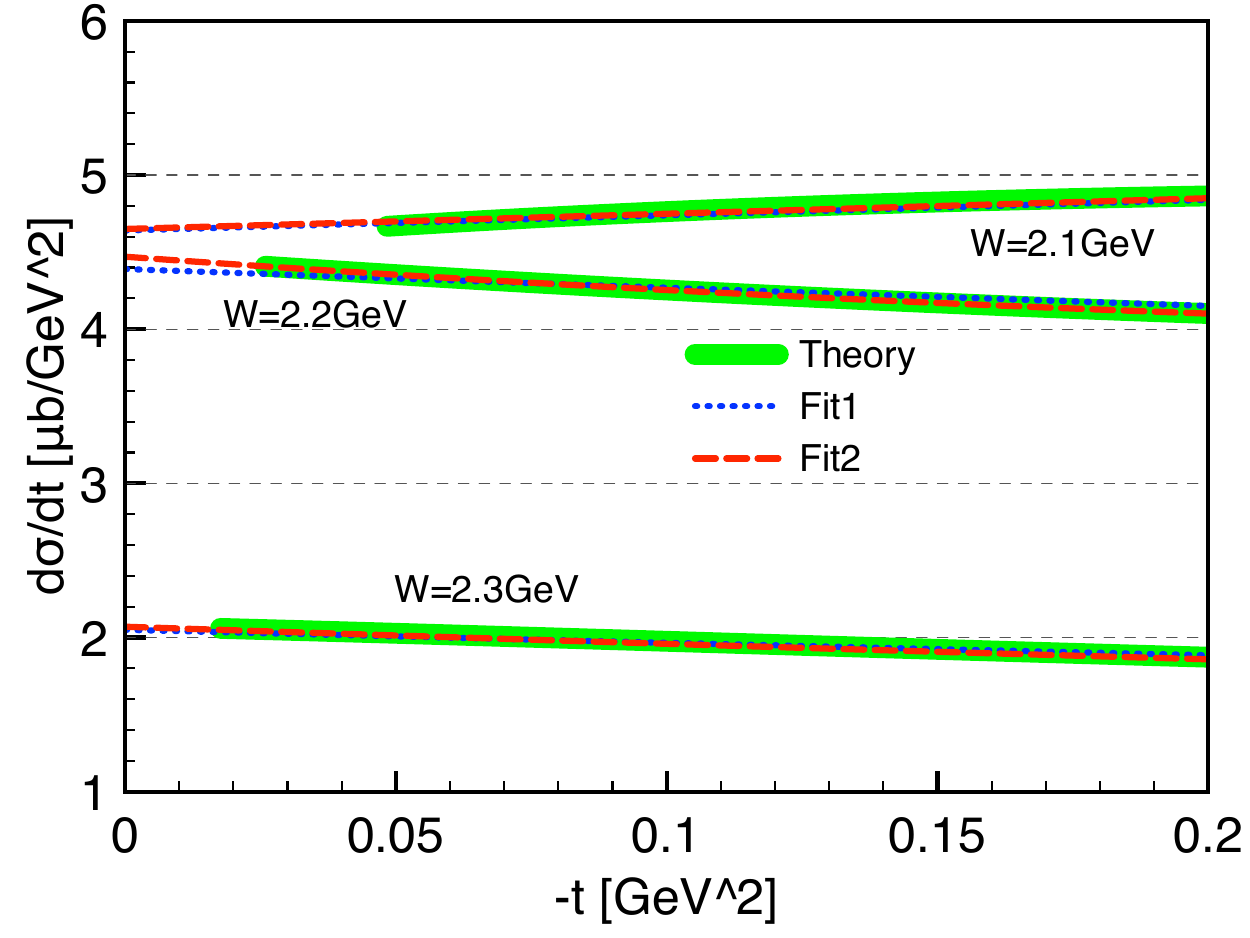}}{-0.3cm}{0.5cm}
\topinset{(b)}{\includegraphics[width= 7.5cm]{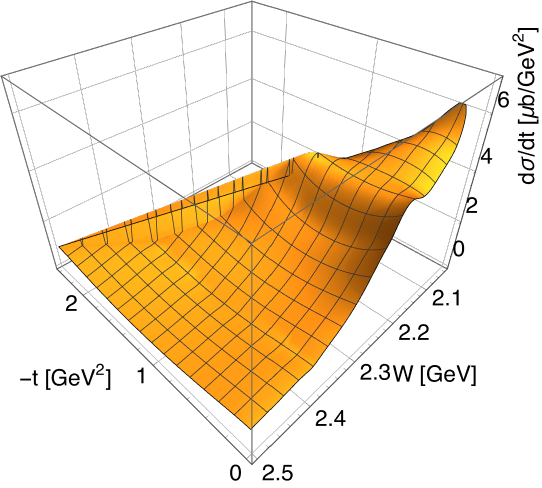}}{-0.3cm}{0.5cm}
\end{tabular}
\caption{(Color online) (a) Forward differential cross-sections $d\sigma/dt$ as functions of $-t$ for different $W$. (b) The same as a function of $-t$ and $W$.}
\label{FIG4}
\end{figure}
In panel (a) of Fig.~\ref{FIG4}, we present the full results for the forward differential cross-sections, $d\sigma/dt$, as functions of $-t$ for different values of $W$. As expected from Fig.~\ref{FIG3}, the curve shapes become more concave with increasing $W$ due to the $N$ contribution. To make the current numerical results more accessible, we separately fit the curves using single-exponential $(d\sigma/dt=ae^{-b|t|})$ and a double-exponential $(d\sigma/dt=a'e^{-b'|t|}+c'e^{-d'|t|})$ functions in the region below $|t|<0.2\rm\,GeV^2$. It is well discussed that the exponent fits indicate the Regge nature of the amplitudes. However, even with the Feynmann propagator as in the present work provides simple exponent fits $d\sigma/dt\propto m^{-4}\exp[-2t/m^2]$ for small $|t|$, where $m$ stands for the mass of the exchange particle in the $t$ channel. We also do not consider the Regge propagator here since 1) our region of interest is confined at relatively low energy, i.e., about $0.5$ GeV above the threshold $W\lesssim2.5$ GeV and 2) the Regge trajectories for $S=1/2$ nucleon exchanges are not fully understood. The corresponding fit parameters are listed in Table~\ref{TAB3}. To better understand the overall $t$-dependence of the cross-section, we plot it as a function of both $-t$ and $W$. A bump structure appears around $W=2.2$ GeV, indicating the contribution from the $f_{0,2}$ mesons.

\begin{table}[b]
\begin{tabular}{|c|c|c|c|c|c|c|} 
\hline
$W$ [GeV]&$a$&$b$&$a'$&$b'$&$c'$&$d'$\\
\hline
$2.1$&$4.64$&$-0.21$&$3.02$&$-0.21$&$1.63$&$-0.21$\\
$2.2$&$4.39$&$0.28$&$2.72$&$-0.38$&$1.75$&$2.03$\\
$2.3$&$2.05$&$0.42$&$2.06$&$0.60$&$0.02$&$-3.50$\\
\hline
\end{tabular}
\caption{Parameters for the single $(d\sigma/dt=ae^{-b|t|})$ and double exponent $(d\sigma/dt=a'e^{-b'|t|}+c'e^{-d'|t|})$ fits. All the parameters are in the $1/\mathrm{GeV}^2$ unit.}
\label{TAB3}
\end{table}

\begin{figure}[t]
\begin{tabular}{ccc}
\topinset{Adair $\rho^{\lambda}_{00}$}{\includegraphics[width= 5.5cm]{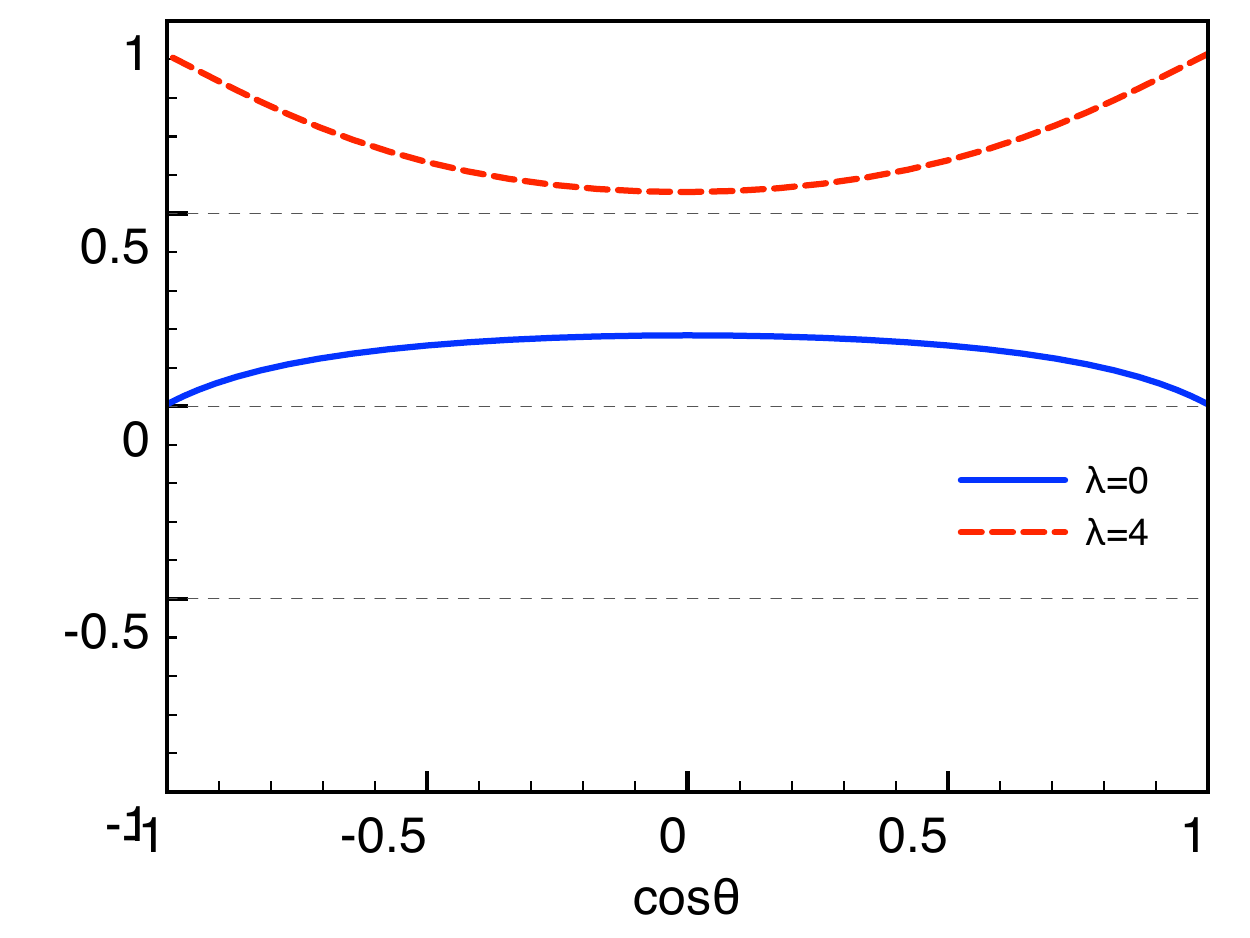}}{-0.4cm}{0.5cm}
\topinset{Adair $\rho^{\lambda}_{10}$}{\includegraphics[width= 5.5cm]{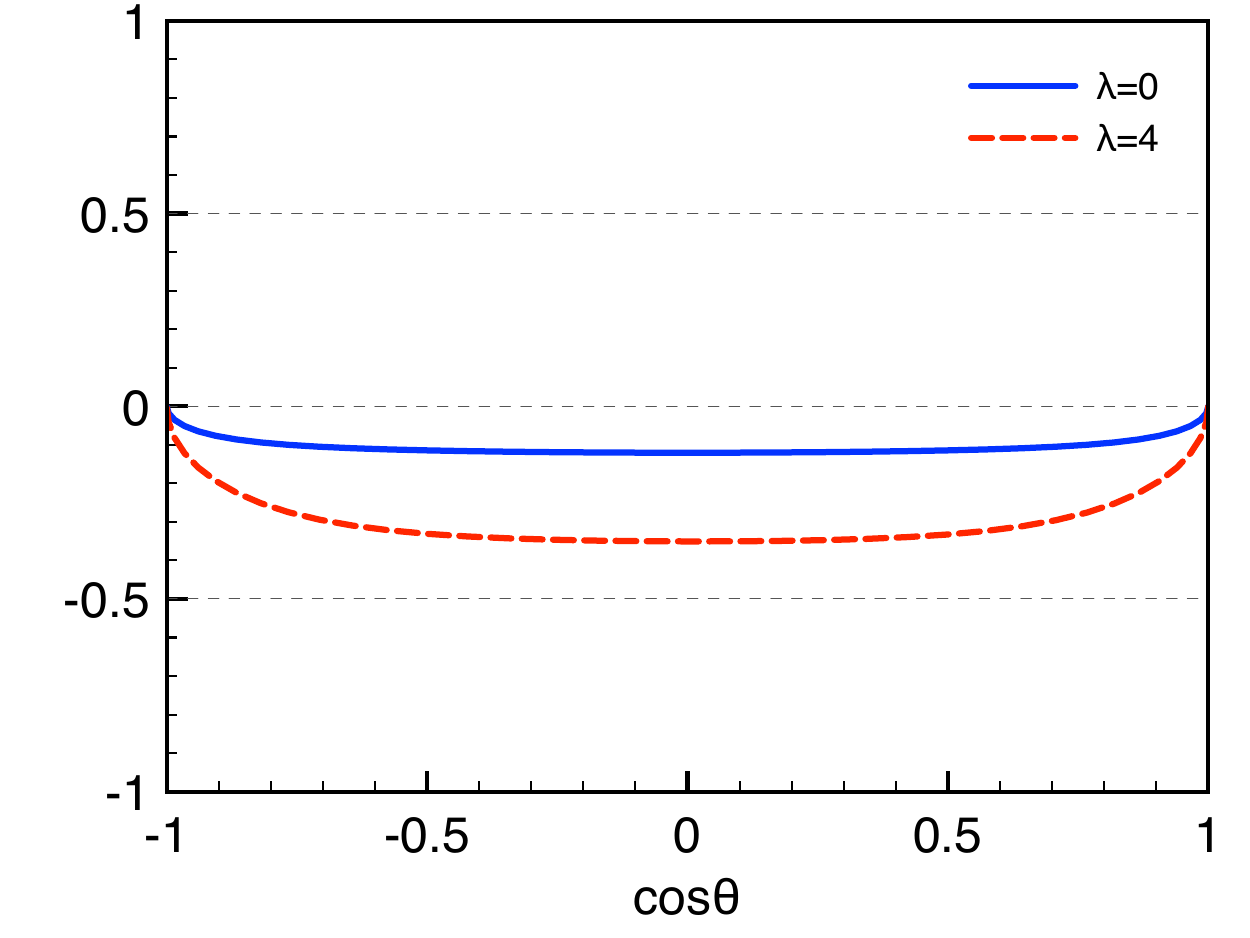}}{-0.4cm}{0.5cm}
\topinset{Adair $\rho^{\lambda}_{1-1}$}{\includegraphics[width= 5.5cm]{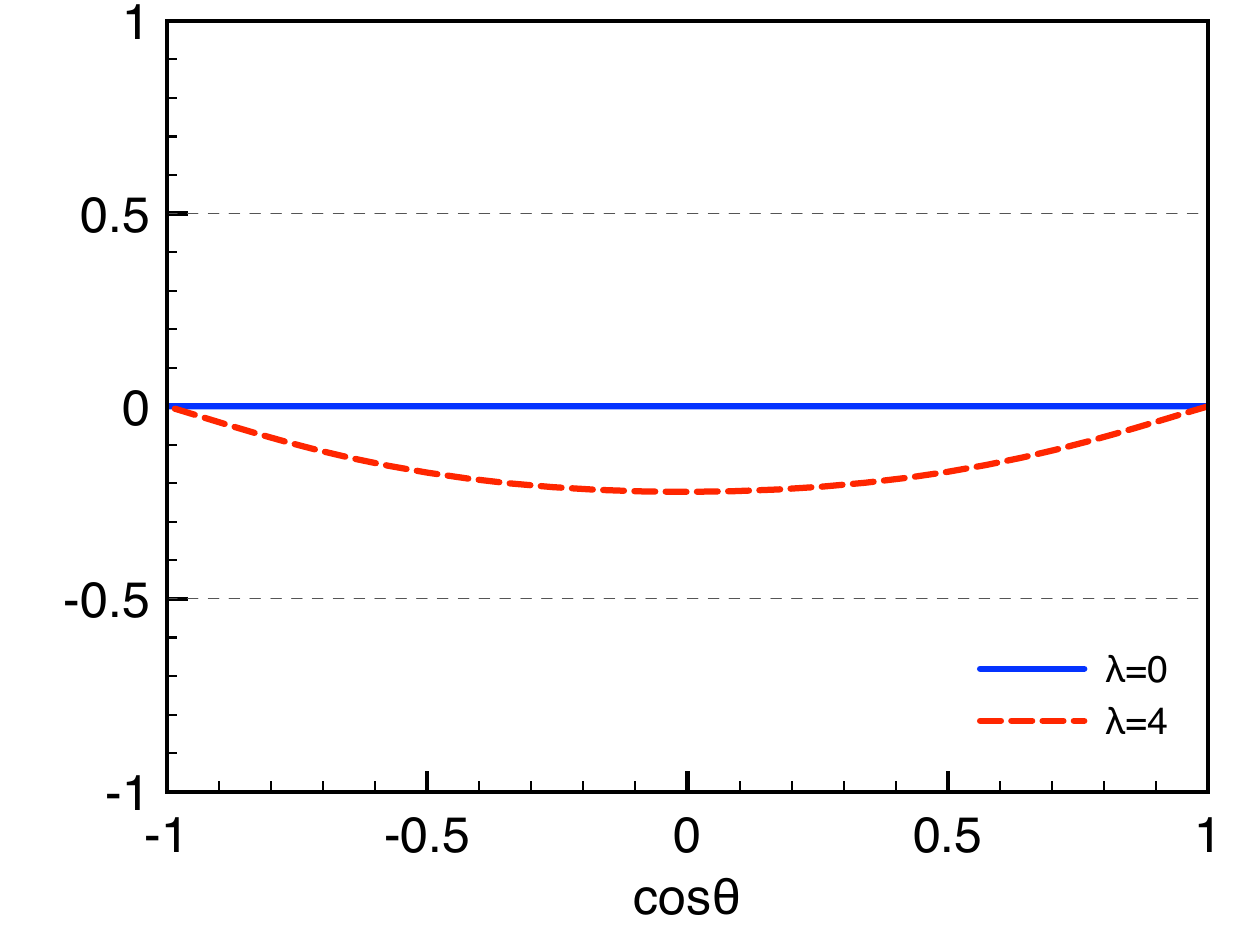}}{-0.4cm}{0.5cm}
\end{tabular}
\begin{tabular}{ccc}
\topinset{Helicity $\rho^{\lambda}_{00}$}{\includegraphics[width= 5.5cm]{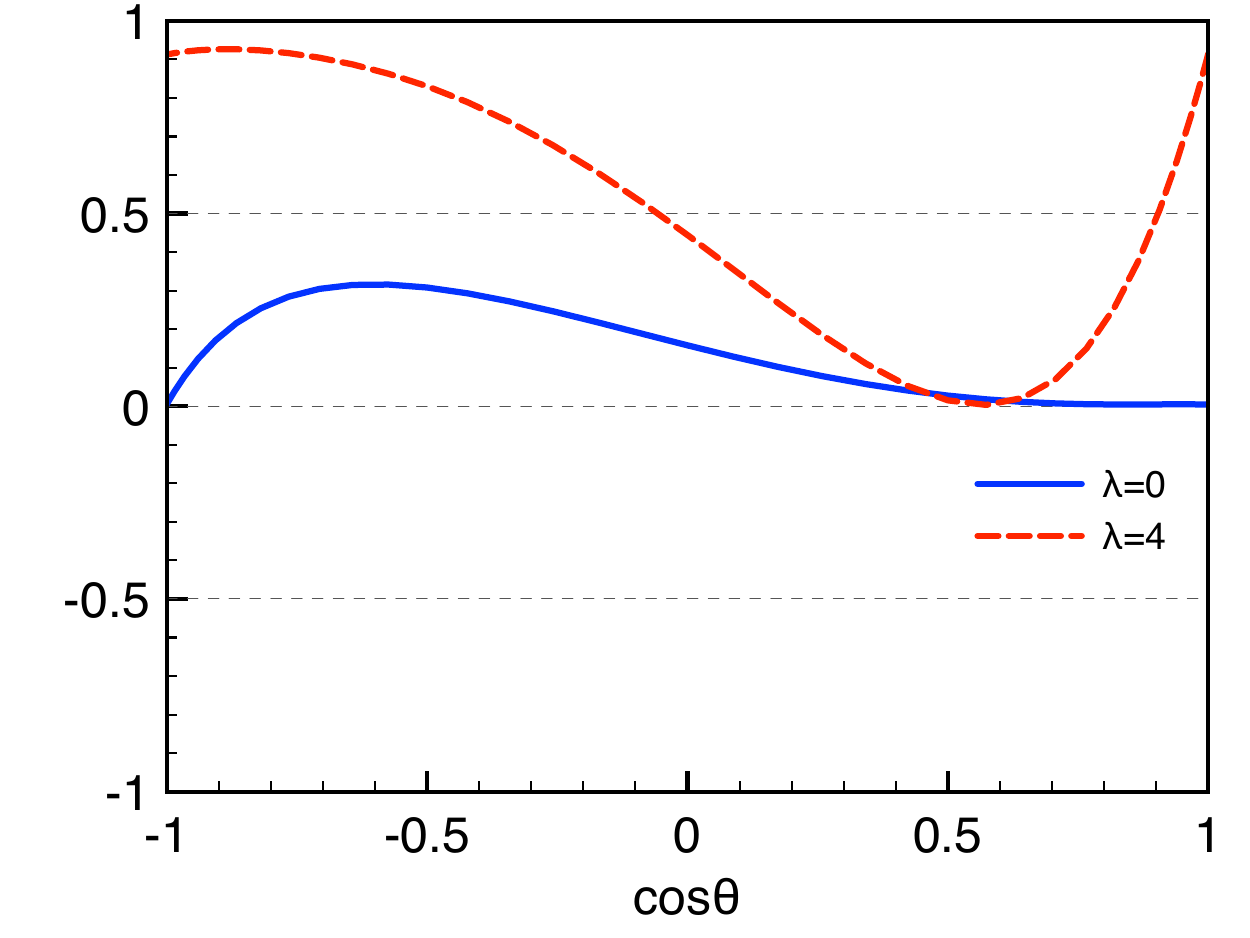}}{-0.4cm}{0.5cm}
\topinset{Helicity $\rho^{\lambda}_{10}$}{\includegraphics[width= 5.5cm]{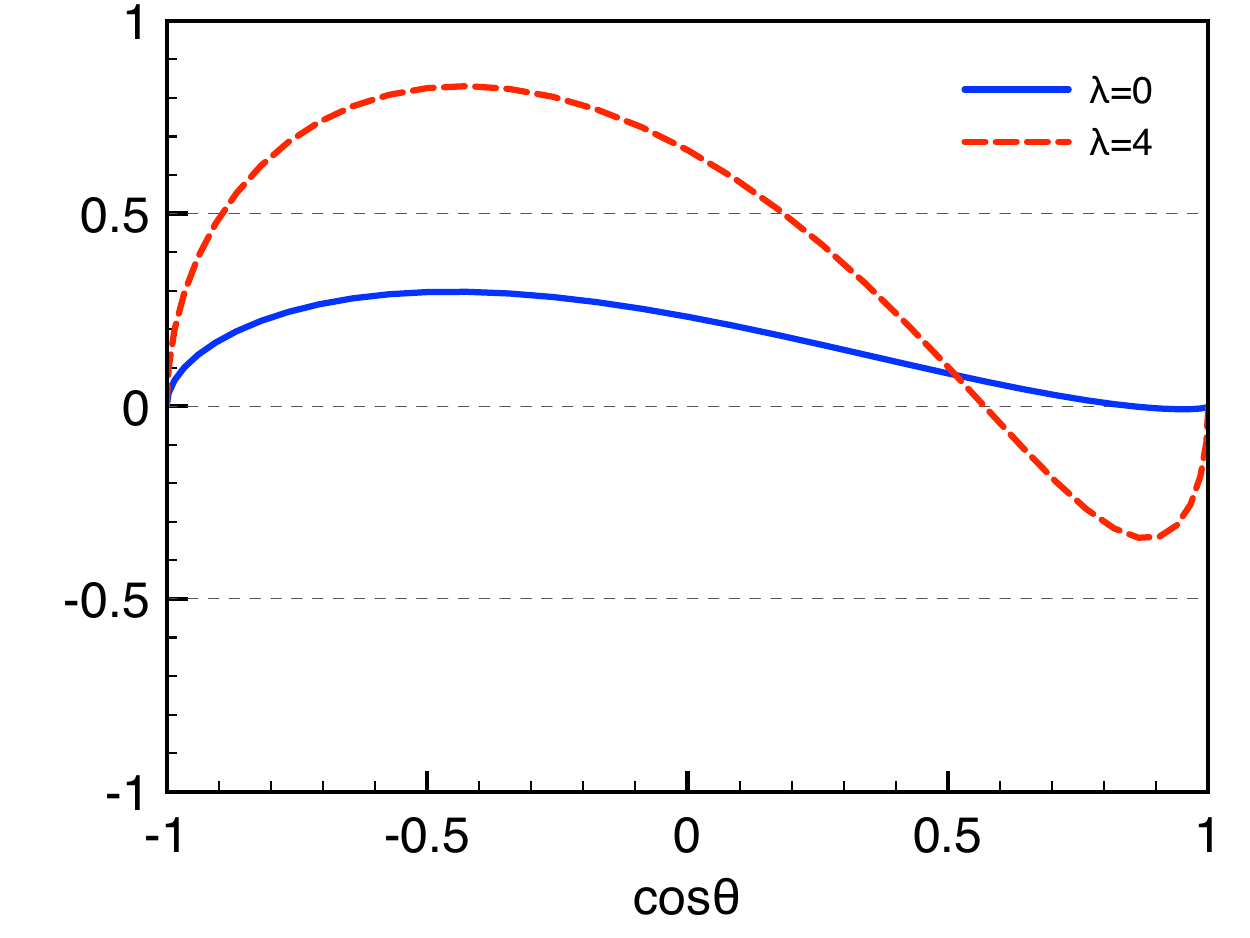}}{-0.4cm}{0.5cm}
\topinset{Helicity $\rho^{\lambda}_{1-1}$}{\includegraphics[width= 5.5cm]{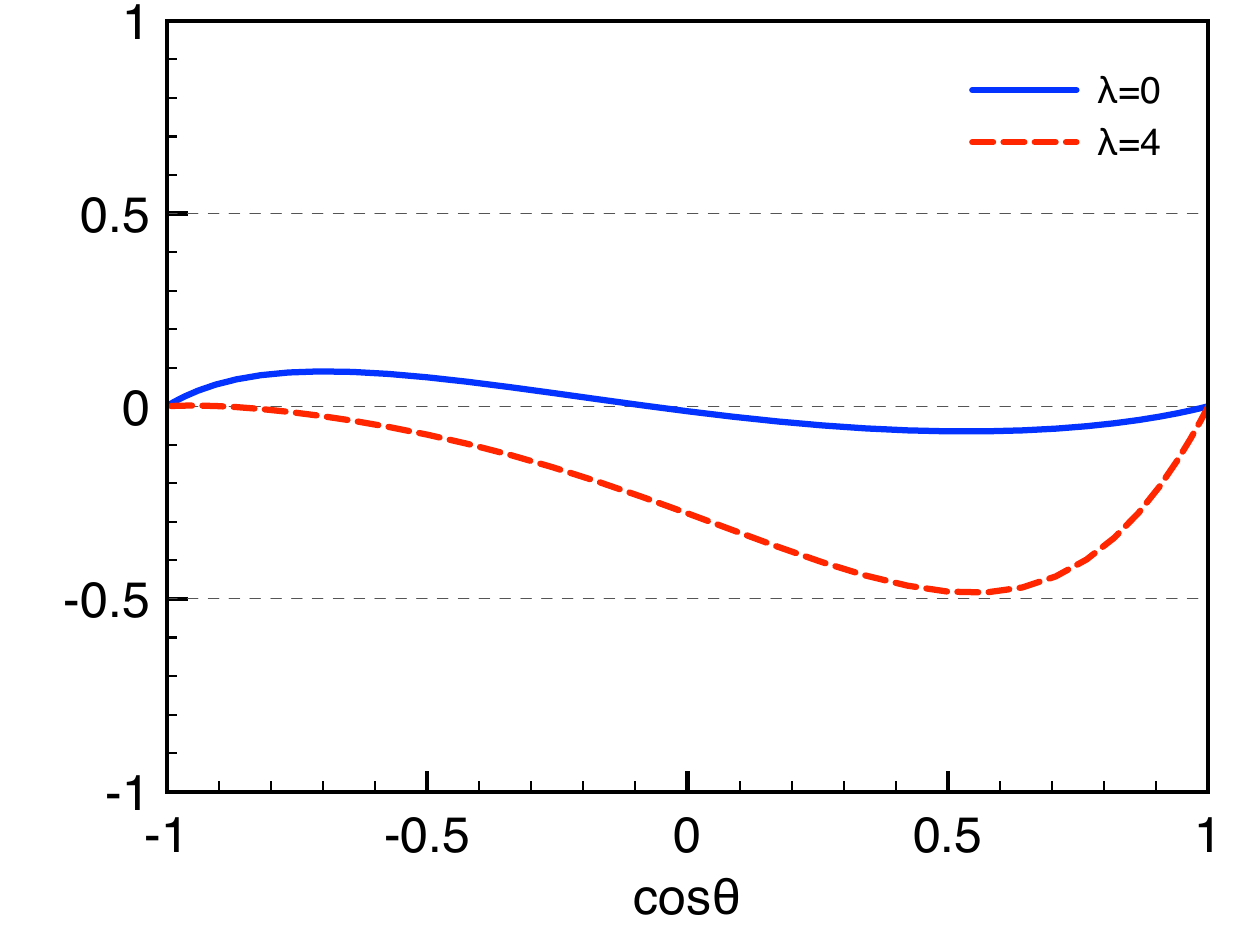}}{-0.4cm}{0.5cm}
\end{tabular}
\begin{tabular}{ccc}
\topinset{GJ $\rho^{\lambda}_{00}$}{\includegraphics[width= 5.5cm]{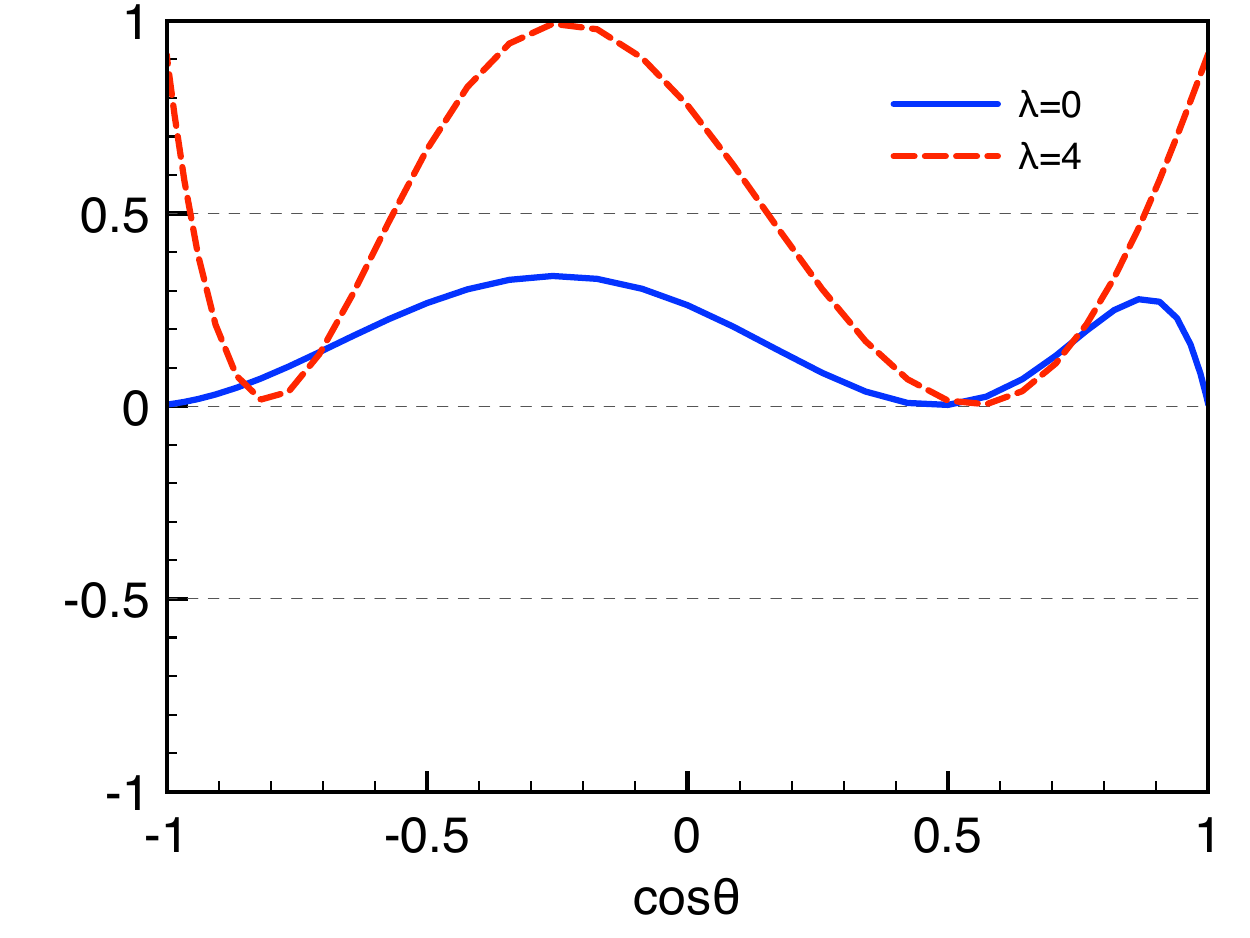}}{-0.4cm}{0.5cm}
\topinset{GJ $\rho^{\lambda}_{10}$}{\includegraphics[width= 5.5cm]{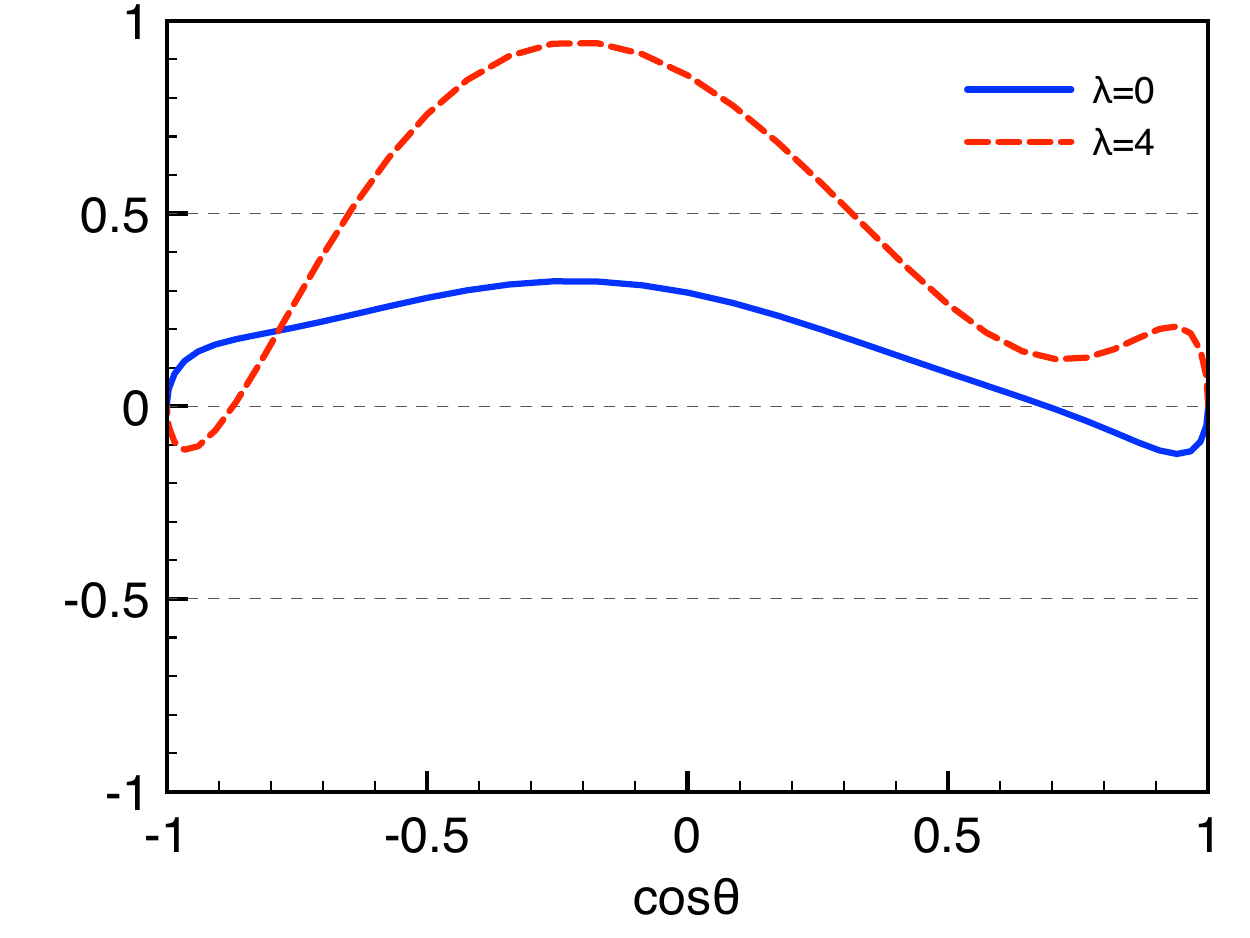}}{-0.4cm}{0.5cm}
\topinset{GJ $\rho^{\lambda}_{1-1}$}{\includegraphics[width= 5.5cm]{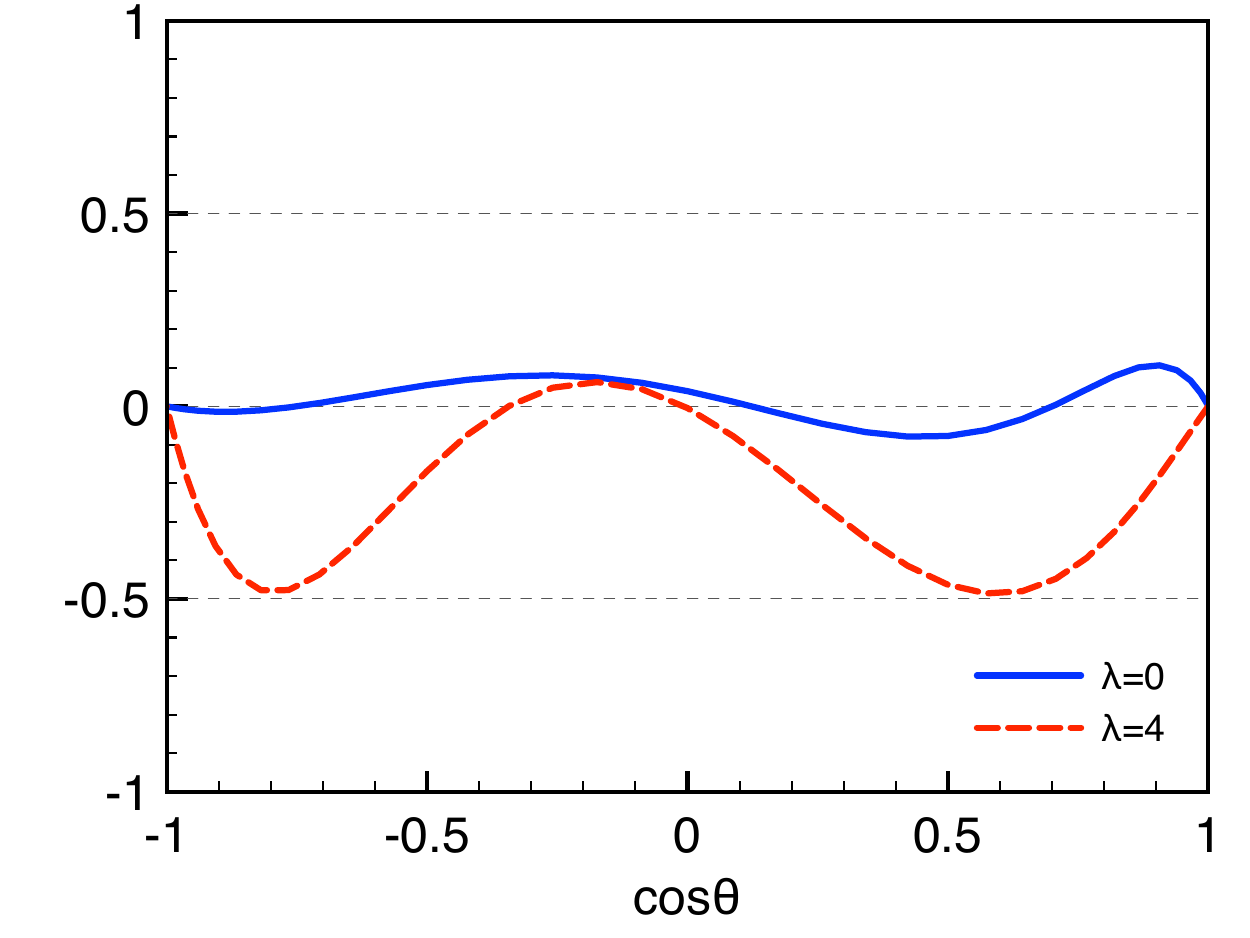}}{-0.4cm}{0.5cm}
\end{tabular}
\caption{(Color online) Spin-density matrix elements (SDMEs) $\rho^\lambda_{00,10,1-1}$ as functions of $\cos\theta$ for the Adair, helicity, and Gottfried-Jackson (GJ) frames for $\lambda=(0,4)$, which stands for the $\phi_4$ helicity ($\pm1$,0), at $W=2.2$ GeV.}
\label{FIG5}
\end{figure}
Now, we turn to the discussion of the spin-density matrix elements (SDMEs) as defined in Eq.~(\ref{eq:SDME}). The numerical results for $\rho^\lambda_{00,10,1-1}$ are plotted in Fig.~\ref{FIG5} as functions of $\cos\theta$ for the full calculation, shown across different kinematic frames, namely, the Adair, helicity, and Gottfried-Jackson (GJ) frames, where $\lambda = (0, 4)$ represents the $\phi_4$ helicity at $W = 2.2$ GeV. According to Eq.~(\ref{eq:SDME}), each SDME approximately follows specific helicity-flip patterns:
\begin{eqnarray}
\label{eq:SDMEX}
&&\rho^0_{00}\propto
|\mathcal{M}_{01}|^2
+|\mathcal{M}_{0-1}|^2,\,\,\,\,\,\,
\rho^4_{00}\propto|\mathcal{M}_{00}|^2,
\cr
&&\rho^0_{10}\propto
\left(\mathcal{M}_{11}
\mathcal{M}^*_{01}
+
\mathcal{M}_{1-1}
\mathcal{M}^*_{0-1}\right),\,\,\,
\rho^4_{10}\propto
\mathcal{M}_{10}
\mathcal{M}^*_{00},
\cr
&&\rho^0_{1-1}\propto\left(
\mathcal{M}_{11}
\mathcal{M}^*_{-11}
+
\mathcal{M}_{1-1}
\mathcal{M}^*_{-1-1}\right),\,\,\,
\rho^4_{1-1}\propto
\mathcal{M}_{10}
\mathcal{M}^*_{-10},
\end{eqnarray}
where the amplitude is defined as $\mathcal{M}_{\lambda_3\lambda_4}$. Here, we define the notation $\Delta\lambda_{34} = |\lambda_3 - \lambda_4|$. From the numerical results shown in Fig.~\ref{FIG5} and being understood by Eq.~(\ref{eq:SDMEX}), we clearly observe that the single $(\Delta\lambda_{34}=1)$ and double $(\Delta\lambda_{34}=2)$ helicity-flip SDMEs become zero at $\cos\theta = \pm 1$, indicating helicity conservation. In contrast, the $\Delta\lambda_{34}=0$ component remains finite~\cite{Kim:2019kef}. We also find that $\rho^4_{00}$ is not exactly unity at $\cos\theta = \pm 1$ due to finite helicity non-conserving effects from the $f_2$ and $N^*$ contributions. As expected, the $\Delta\lambda_{34}=2$ contribution is very small, as seen from $\rho^0_{1-1}$. Notably, the shape of the SDMEs is primarily driven by the $N$ contribution, which plays a dominant role in the total cross-section. 

\begin{figure}[t]
\begin{tabular}{ccc}
\topinset{Adair}{\includegraphics[width=5.5cm]{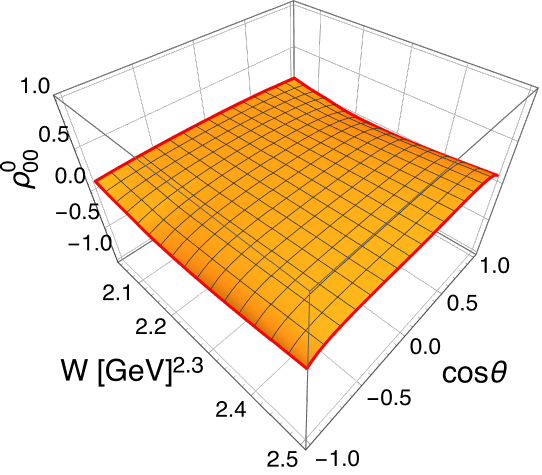}}{-0.4cm}{0.5cm}
\topinset{Helicity}{\includegraphics[width=5.5cm]{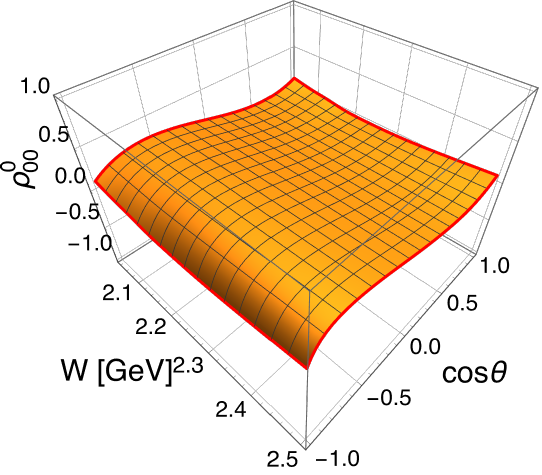}}{-0.4cm}{0.5cm}
\topinset{GJ}{\includegraphics[width=5.5cm]{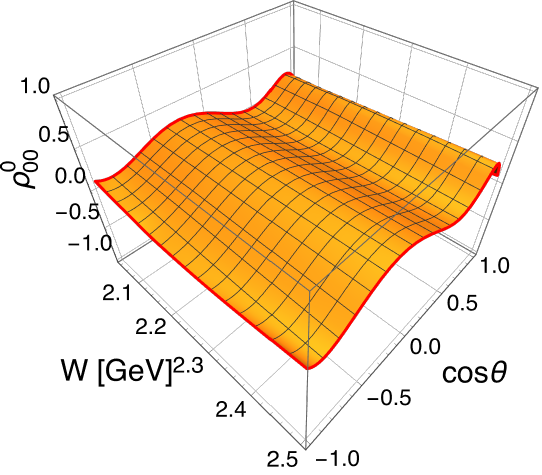}}{-0.4cm}{0.5cm}
\end{tabular}
\begin{tabular}{ccc}
\topinset{Adair}{\includegraphics[width=5.5cm]{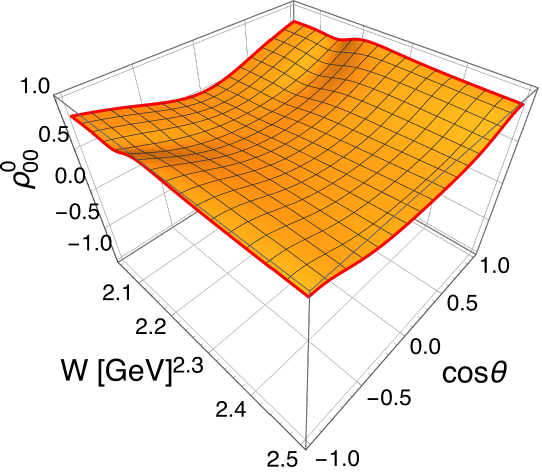}}{-0.4cm}{0.5cm}
\topinset{Helicity}{\includegraphics[width=5.5cm]{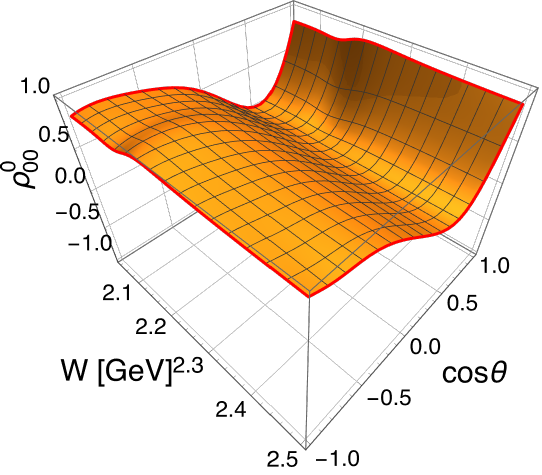}}{-0.4cm}{0.5cm}
\topinset{GJ}{\includegraphics[width=5.5cm]{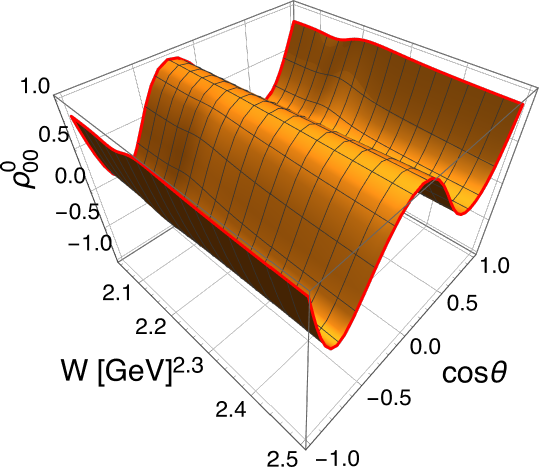}}{-0.4cm}{0.5cm}
\end{tabular}
\caption{(Color online) Spin-density matrix elements (SDMEs) $\rho^{0,4}_{00}$ as functions of $W$ and $\cos\theta$ for the Adair, helicity, and Gottfried-Jackson (GJ) frames.}
\label{FIG6}
\end{figure}

In Fig.~\ref{FIG6}, we plot $\rho^{0,4}_{00}$ as functions of $W$ and $\cos\theta$ for the Adair, helicity, and Gottfried-Jackson (GJ) frames. The energy dependence of the SDMEs is shown to be quite mild. At the same time, meson contributions, such as from the $f_0$, introduce a small but non-trivial structure around $W = 2.2$ GeV for helicity-conserving cases, i.e., $\lambda = 4$ ($\Delta\lambda_{34} = 0$). As expected, the $\Delta\lambda_{34} = 0$ contributions are significantly larger than those with $\Delta\lambda_{34} = 1$.

\begin{figure}[b]
\begin{tabular}{cc}
\topinset{(a)}{\includegraphics[width= 8.5cm]{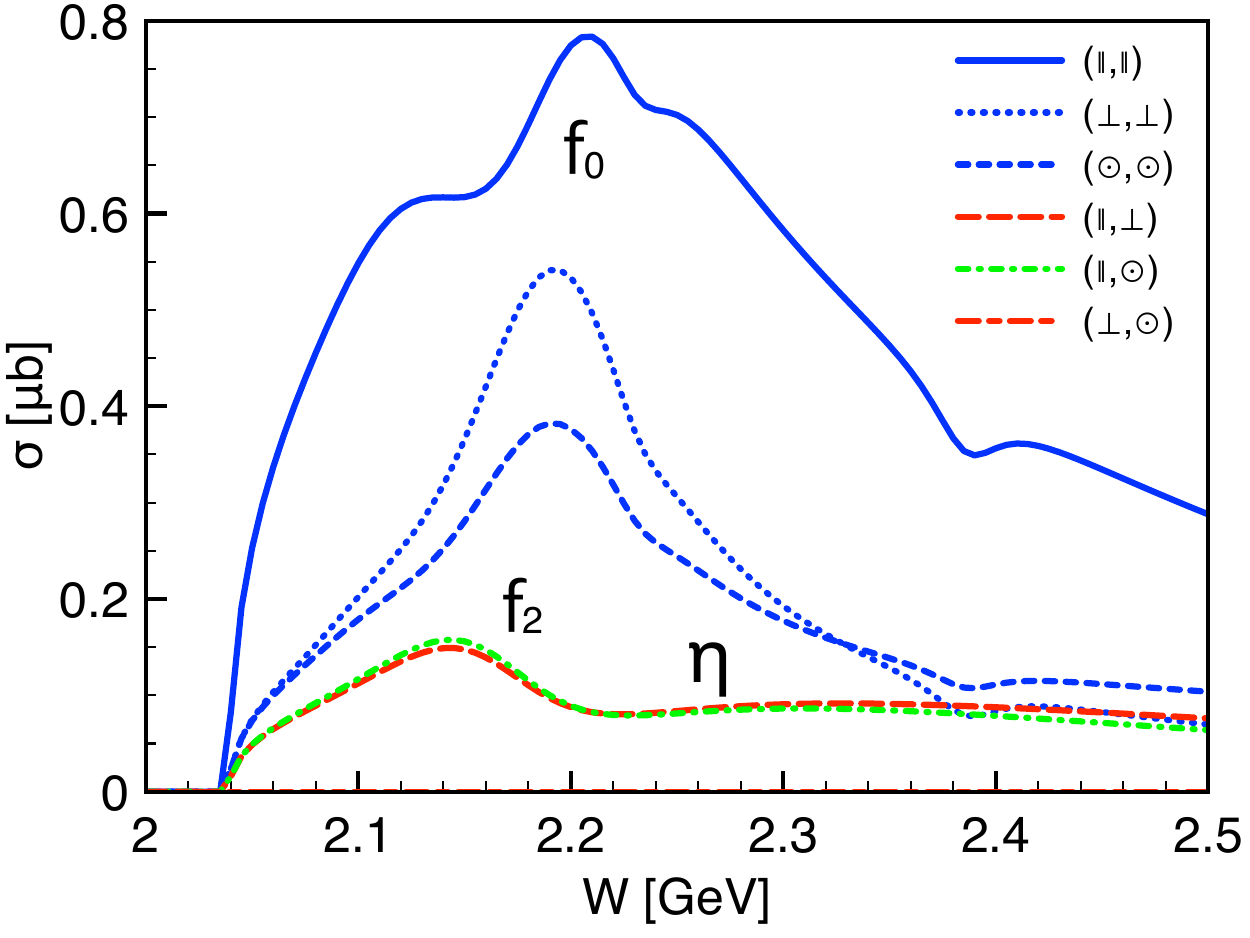}}{-0.3cm}{0.5cm}
\topinset{(b)}{\includegraphics[width= 8.5cm]{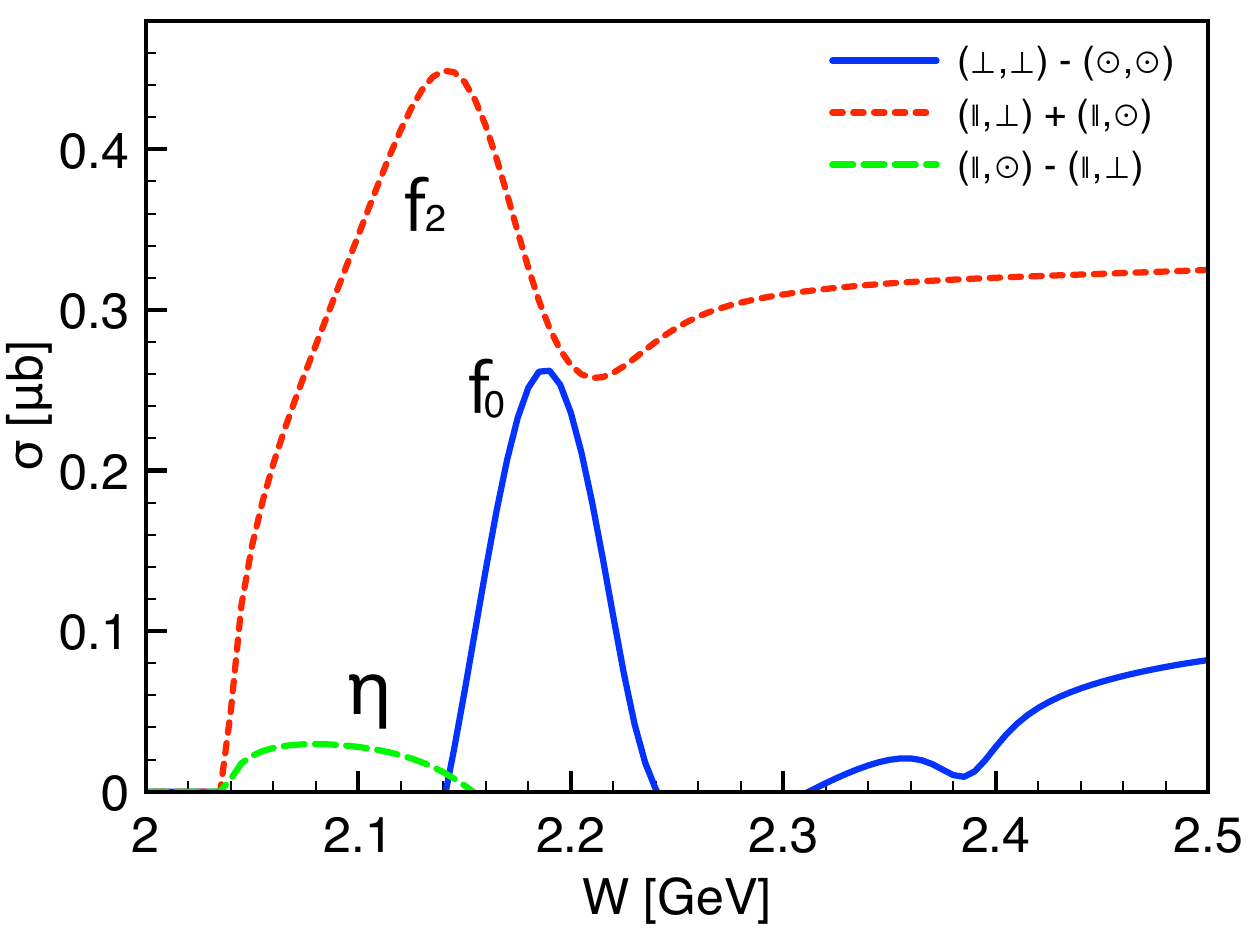}}{-0.3cm}{0.5cm}
\end{tabular}
\begin{tabular}{cc}
\topinset{(c)}{\includegraphics[width= 8.5cm]{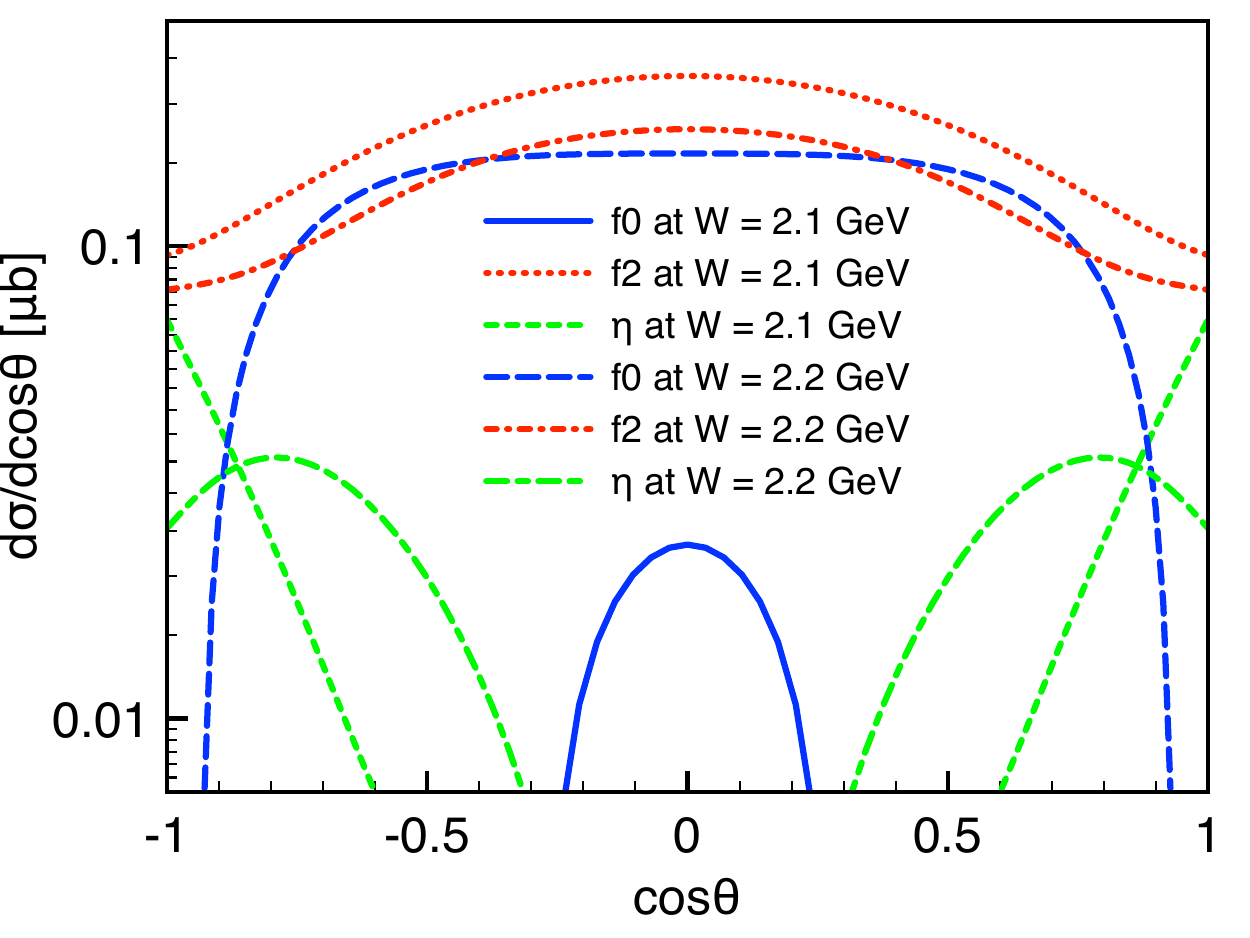}}{-0.3cm}{0.5cm}
\topinset{(d)}{\includegraphics[width= 8.5cm]{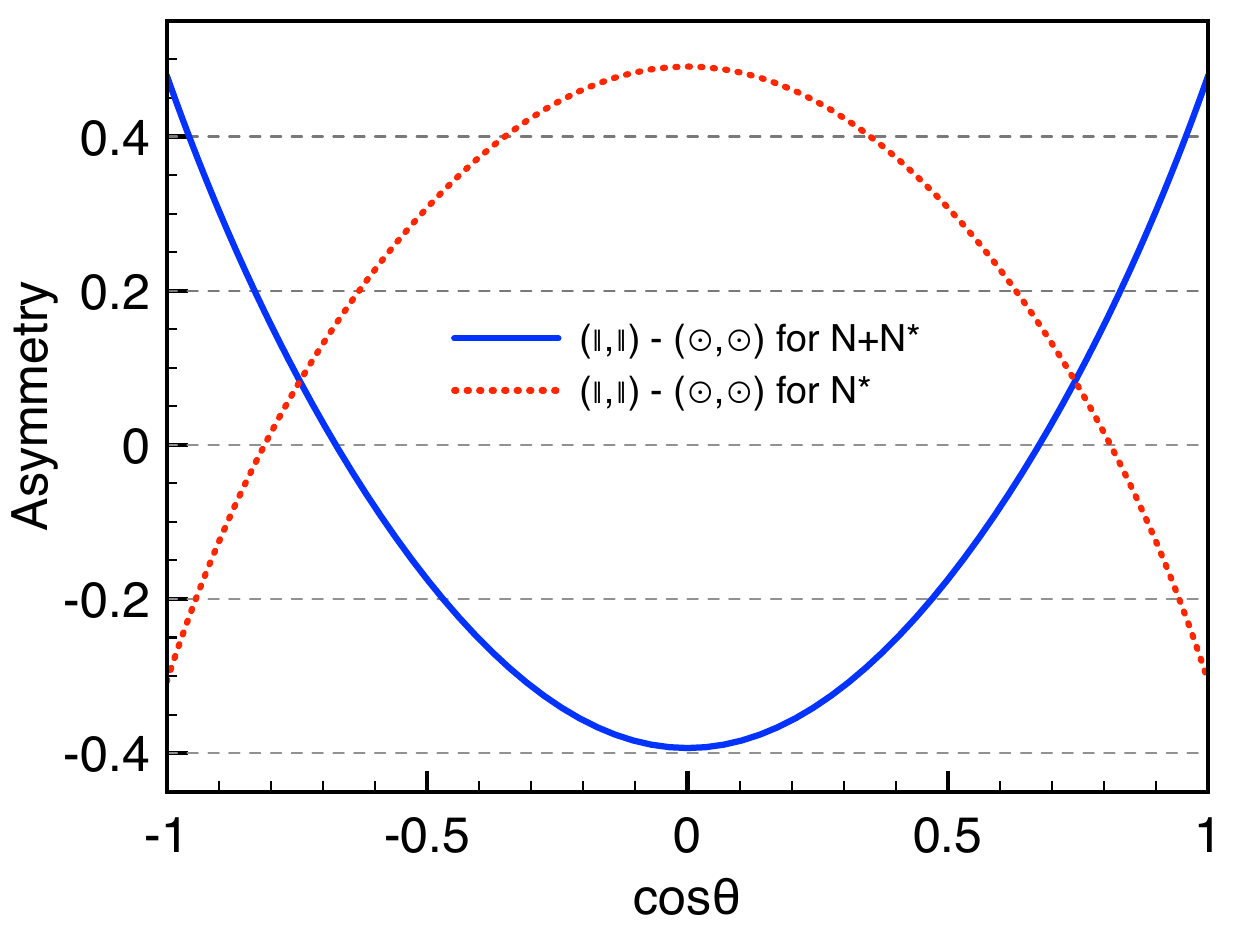}}{-0.3cm}{0.5cm}
\end{tabular}
\caption{(Color online) (a) Polarized total cross-sections as functions of $W$ for the different combinations of $\phi_3$ and $\phi_4$ polarizations, i.e., $(\epsilon_{\phi_3},\epsilon_{\phi_4})$. (b) Added and subtracted polarized cross-sections for the different polarization combinations to enhance the meson signals. (c) Polarized differential cross-sections as functions of $\cos\theta$ in the same manner as the panel (b) for $W=(2.1-2.3)$ GeV. (d) Polarization asymmetries in Eq.~(\ref{eq:PA}) as functions of $\cos\theta$ with $N+N^*$ and $N^*$.}
\label{FIG7}
\end{figure}

Finally, we turn to the discussion of polarization observables, which can provide valuable insight into reaction mechanisms by examining various combinations of $\phi$-meson polarizations. In panel (a) of Fig.~\ref{FIG7}, we present the numerical results for polarized total cross-sections as functions of $W$ for different combinations of $\phi_3$ and $\phi_4$ polarizations, denoted as $(\epsilon_{\phi_3},\epsilon_{\phi_4})$. The symbols $\perp$ and $\parallel$ indicate that the polarizations are, respectively, transverse and parallel to the reaction plane, while $\odot$ denotes the longitudinal polarization. It is evident from the Lorentz structure of the invariant amplitudes in Eq.~(\ref{eq:AMP}), it is clear that the amplitudes are sensitive to polarization and are reduced by certain combinations.

The polarized total cross-sections show significant contributions from $N$ and $N^*$ for identical polarization combinations, but these contributions decrease for different combinations. As described by Eq.~(\ref{eq:AMP}), the $f_0$ amplitude becomes zero for the combinations $(\parallel,\perp)$ and $(\parallel,\odot)$, whereas the $\eta$ amplitude remains non-zero only for $(\parallel,\odot)$. In contrast, the $f_2$ amplitude remains finite for both combinations. This pattern is illustrated in panel (a) of Fig.~\ref{FIG7}, showing the bumps corresponding to $f_{0,2}$ and $\eta$, which suggests that meson signals can be enhanced by appropriately adding or subtracting the contributions from different polarization combinations. This approach is tested in panel (b) of Fig.~\ref{FIG7}, where the $f_0$ and $f_2$ contributions are more pronounced and better separated due to improved signal-to-background ratios. In panel (c), we present the angular-dependent differential cross-sections in the same manner as in panel (b) for $W = (2.1 - 2.3)$ GeV. The $f_2$ and $\eta$ contributions exhibit qualitatively flat curves near zero degrees, while the $f_0$ component shows distinctive angular dependence. Analyzing these polarized angular dependencies allows one to isolate and study specific meson properties more effectively.

As mentioned previously, Ref.~\cite{Shi:2010un,Xie:2014tra,Xie:2007qt} considered that the $N^*(1530)$ dominates the background of the present reaction process, whereas we include more $N^*$ and $N$ contributions. To test these two different scenarios, we suggest measuring an asymmetry characterized by the various combinations of the $\phi$-meson polarizations as follows:
\begin{eqnarray}
\label{eq:PA}
\mathrm{Asymmerty}\equiv\frac{d\sigma_{\epsilon_3\epsilon_4}-d\sigma_{\epsilon'_3\epsilon'_4}}
{d\sigma_{\epsilon_3\epsilon_4}+d\sigma_{\epsilon'_3\epsilon'_4}},
\end{eqnarray}
where the $\phi$-meson polarizations are given by $\epsilon_{3,4}=(\parallel,\perp,\odot)$ and $d\sigma\equiv d\sigma/d\cos\theta$. In panel (d) of Fig.~\ref{FIG7}, we show the polarization asymmetries in Eq.~(\ref{eq:PA}) as functions of $\cos\theta$ for the full calculation with $N+N^*$ and that with $N^*$ only, respectively. As seen in panel (a) of Fig.~\ref{FIG7}, these two polarization combinations contain much information on the baryon exchanges. It turns out that the angular dependences are distinctively different for the two cases, especially at $\cos\theta=\pm1$ and $\cos\theta\approx0$, and these differences can be tested in experiments to pin down a reaction mechanism. 

\section{Summary and future perspectives}
The present study investigated the production process of two $\phi$ mesons in antiproton-proton annihilation, near the threshold energy. As discussed, the process is notable for exhibiting a substantial violation of the Okubo-Zweig-Iizuka (OZI) rule. We find that the phenomenological hadronic process can describe the available data without quark and gluon degrees of freedom, such as the glueball, indicating the importance of the vector-meson--baryon coupled-channel interactions, which are implicitly included in the phenomenological strong couplings. This theoretical investigation also offers predictions for upcoming experiments, such as E104~\cite{Ahn:2024} planned at J-PARC, which will measure polarization observables for the first time in this context.

We use an effective Lagrangian approach to study the present reaction, incorporating multiple interaction channels. In the $t$ and $u$ channels, the exchanges of the ground-state nucleon and its excited states $N^*(1535,1650,1895,2071)$ are considered, which have significant couplings to the $\phi N$ channel. The pentaquark-like $P_s$ is also taken into account. The $s$ channel includes the scalar $f_0(2020,2100,2200)$ and tensor $f_2(1950,2010,2150)$ mesons, as well as the pseudoscalar $\eta(2225)$. These exchanges allow the study to capture key dynamics responsible for the observed cross-sections and polarization patterns.

The numerical results show several interesting phenomena. The nucleon and various resonances dominate near the reaction threshold, rapidly increasing the total cross-section. The cusp structures appear near the $\bar{\Lambda}\Lambda$ and $\bar{\Sigma}\Sigma$ thresholds, affecting the behavior of the nontrivial structures of the cross-section. In addition, the scalar and tensor mesons like $f_0$ and $f_2$ introduce distinct peaks around $2.2$ GeV in the cross-section. These contributions highlight the interplay between baryon and meson contributions, challenging previous theoretical models considering only the $N^*(1535)$ resonance.

The angular dependence of the cross-section turns out to be symmetric for $\theta=\pi/2$, due to the identical final-state particles, and the nucleon $t$-channel exchange dominates its shape. The polarization observables were also analyzed using spin-density matrix elements (SDMEs), which reveal detailed patterns in the scattering angles and helicity conservation and show that helicity-conserving components are dominant. These polarization patterns provide crucial information about the reaction mechanism and can serve as benchmarks for future experiments.

The study concludes that a combination of meson and baryon contributions can explain the observed violation of the OZI rule in the present reactions. The nucleon resonances and scalar/tensor mesons contribute significantly to the near-threshold dynamics. The presence of the $\bar{\Lambda}\Lambda$ cusp and the complex polarization observables further support the need for a more detailed experimental investigation. Upcoming experiments at J-PARC will be essential for validating the theoretical predictions presented in this study.

In summary, this research provides new insights into the mechanisms behind double $\phi$ meson production, highlighting the importance of scalar and tensor mesons and violation of the OZI rule in QCD regarding the hadronic degrees of freedom. The detailed analysis of cross-sections, polarization observables, and resonance contributions offers a comprehensive framework for understanding this process, laying the groundwork for future experimental work. We emphasize that further experiments will be necessary to confirm the theoretical findings, particularly those concerning polarization and the role of excited nucleon and meson states. Theoretically, coupled-channel calculations are also a proper way to investigate the present reaction process, exploring the violation of the OZI rule directly~\cite{KB}. Related works are currently in progress and will appear elsewhere. 

\section*{Acknowledgment}
The authors are grateful for the insightful discussions with Sang-Ho Kim (Soongsil University),  Kanchan Pradeepkumar Khemchandani (Federal University of São Paulo), and Alberto Martinez Torres (University of São Paulo). The work of S.i.N. and D.Y.L. was supported by grants from the National Research Foundation of Korea (NRF), funded by the Korean government (MSIT) (NRF-2022R1A2C1003964, RS-2021-NR060129, and RS-2024-00436392).  The work of D.Y.L. was also supported partially by Global-Learning \& Academic research institution for Master’s$\cdot$PhD students, and Postdocs (G-LAMP) Program of the National Research Foundation of Korea (NRF) grant funded by the Ministry of Education (RS-2023-00301702).

\end{document}